\documentclass{svmult}

\usepackage{graphicx}			
\usepackage{fancyhdr}			
\usepackage{geometry}			
\usepackage{amsmath}			
\usepackage{amssymb}
\usepackage{latexsym}
\usepackage{subcaption} 
\usepackage{float} 
\usepackage{braket}
\usepackage{bbold} 
\usepackage{mathtools}
\usepackage{physics}
\usepackage{dsfont}
\usepackage{tikz-feynman}
\usepackage{listings}
\usepackage{bm}

\usepackage{mcite}
\usepackage{cite}

\usepackage[hyperfootnotes=false,colorlinks=false]{hyperref}
\hypersetup{
    colorlinks,%
    citecolor=black,%
    filecolor=black,%
    linkcolor=black,%
    urlcolor=cyan,
    }
\usepackage{url}
\usepackage{color}
\usepackage{setspace}
\usepackage{inputenc}
\usepackage[english]{babel}

\usepackage{xcolor}

\DeclareFontFamily{U}{mathb}{\hyphenchar\font45}
\DeclareFontShape{U}{mathb}{m}{n}{
      <5> <6> <7> <8> <9> <10> gen * mathb
      <10.95> mathb10 <12> <14.4> <17.28> <20.74> <24.88> mathb12
      }{}
\DeclareSymbolFont{mathb}{U}{mathb}{m}{n}
\DeclareFontSubstitution{U}{mathb}{m}{n}

\let\dot\relax
\DeclareMathAccent{\dot}{0}{mathb}{"39}
\let\ddot\relax
\DeclareMathAccent{\ddot}{0}{mathb}{"3A}
\let\dddot\relax
\DeclareMathAccent{\dddot}{0}{mathb}{"3B}
\let\ddddot\relax
\DeclareMathAccent{\ddddot}{0}{mathb}{"3C}

\newcommand{\id}{\mathbb{1}}

\def\con{\color{black}}

\title*{Classical and Quantum Nonlocal Gravity}

\author{Arnau Bas i Beneito, Gianluca Calcagni\thanks{corresponding author} and Les{\l}aw Rachwa\l{}}
\institute{Arnau Bas i Beneito \at Institut de Física Corpuscular (IFIC), Carrer del Catedràtic José Beltrán Martinez, 2,\\  46980 Paterna, Valencia, Spain\\ \email{Arnau.Bas@ific.uv.es} \and
Gianluca Calcagni \at Instituto de Estructura de la Materia, CSIC,\\ Serrano 121, 28006 Madrid, Spain\\ \email{g.calcagni@csic.es} \and Les{\l}aw Rachwa\l{} \at Departamento de Física -- Instituto de Ciências Exatas, Universidade Federal de Juiz de Fora,\\ 33036-900, Juiz de Fora, MG, Brazil\\ \email{grzerach@gmail.com}}

\begin{document}

\maketitle

\abstract{
This chapter of the Handbook of Quantum Gravity aims to illustrate how nonlocality can be implemented in field theories, as well as the manner it solves fundamental difficulties of gravitational theories. We review Stelle's quadratic gravity, which achieves multiplicative renormalizability successfully to remove quantum divergences by modifying the Einstein's action but at the price of breaking the unitarity of the theory and introducing Ostrogradski's ghosts. Utilizing nonlocal operators, one is able not only to make the theory renormalizable, but also to get rid of these ghost modes that arise from higher derivatives. We start this analysis by reviewing the classical scalar field theory and highlighting how to deal with this new kind of nonlocal operators. Subsequently, we generalize these results to classical nonlocal gravity and, via the equations of motion, we derive significant results about the stable vacuum solutions of the theory. Furthermore, we discuss the way nonlocality could potentially solve the singularity problem of Einstein's gravity. In the final part, we examine how nonlocality induced by exponential and asymptotically polynomial form factors preserves unitarity and improves the renormalizability of the theory.
\keywords{Models of quantum gravity, nonlocality, field theory, classical theories of gravity, renormalization and regularization}}

\newpage

\section*{Conventions}

Throughout this chapter, we will employ the following conventions:
\begin{itemize}
    \item Minkowski spacetime with mostly plus signature, i.e., $\eta_{\mu \nu} = \text{diag}(-1,+1,+1,+1)_{\mu \nu}$.
    \item Einstein summation convention, i.e., $ \sum_i a_i b^i \equiv a_ib^i$.
    \item Greek indices run over the 4 spacetime dimensions whereas Roman indices run only over the 3 spatial dimensions.
    \item We work in natural units, i.e., $\hbar = 1$ and $c = 1$.
    \item The Laplace--Beltrami operator or d'Alembertian $\Box$ is defined as $\Box = \nabla^{\mu}\nabla_{\mu}$.
    \item The gravitational coupling $\kappa$ is defined as $\kappa^2 = 8 \pi G$, where $G$ is Newton's constant.
    \item $A_{[\mu}B_{\nu]}$ denotes anti-symmetrization, i.e., $A_{[\mu}B_{\nu]} = \frac{1}{2}(A_{\mu}B_{\nu}-A_{\nu}B_{\mu})$.
    \item $A_{(\mu}B_{\nu)}$ denotes symmetrization, i.e., $A_{(\mu}B_{\nu)} = \frac{1}{2}(A_{\mu}B_{\nu}+A_{\nu}B_{\mu})$.
    \item The Riemann tensor is defined as $R^{\rho}_{\mu \nu \sigma} = 2 \partial_{[\nu}\Gamma^{\rho}_{ \sigma ]\mu}+ 2 \Gamma^{\lambda}_{\mu [\nu} \Gamma^{\rho}_{\sigma] \lambda}$.
    \item The Ricci tensor is defined as $R_{\mu \nu} = R^{\alpha}_{\mu \alpha \nu}$.
    \item The Ricci scalar is defined as $R = g^{\mu \nu}R_{\mu \nu}$.
    \item The Einstein tensor is defined as $G_{\mu \nu} = R_{\mu \nu} - \frac{1}{2}R g_{\mu \nu}$.
    \item To simplify the index notation, we will refer to the curvature operators $R_{\mu \nu \rho \sigma}$, $R_{\mu \nu}$, and $R$ as $\mathcal{R}$.
\end{itemize}

 
\section{Introduction}

In the early XX century, there occurred two important revolutions in physics. On the one hand, Albert Einstein published what he called the theory of general relativity (GR), in which he was able to describe all the observable gravitational effects through the action
  \begin{equation}
     S = \frac{1}{2\kappa^2}\int d^4x \sqrt{-g}R \label{EHAction} \: .
 \end{equation}
For a local observer, GR recovered special relativity, which Einstein published in 1905 and, together with the posterior geometrical formalism of Hermann Minkowski, introduced the notion of spacetime. This theory was also able to account for new phenomena unexplained by Newtonian gravity, such as the perihelion precession of Mercury, the gravitational time dilation, and the time delay of the light traveling near a massive object \cite{Shapiro:1964uw}. However, along with this novel theory there appeared mathematical predictions for new astronomical objects in the universe called black holes, that opened new mysteries that GR was unable to solve, such as the \emph{singularity problem}, that is, the impossibility of the theory to make predictions in some regions of spacetime where the curvature diverges.

On the other hand, coevally to this theory of gravity, it was discovered that quantum mechanics could successfully describe the microscopic world. From the 1920s on, prominent physicists such as Dirac, Heisenberg, Pauli and later Yukawa, Feynman, Schwinger, Dyson and others developed the formalism of quantum field theory (QFT) within the second quantization prescription and the path-integral formalism. Soon it was shown the great precision this new theory could achieve to compute observable quantities in particle scattering processes.

After the success of the theory of quantum electrodynamics (QED) in the 1940s and 1950s and having been established a solid classical theory of gravity, there appeared several attempts trying to quantize the gravitational theory in the same way done for the other fundamental forces of Nature. However, the complications arising from the very definition of this rank-2 symmetric field theory and its direct correspondence with geometry impeded its realization for decades. During the second half of the XX century and until nowadays, some approaches to this problem of quantum gravity were proposed, such as supergravity, string theory, loop quantum gravity, and others. To various degrees, each of them is well-formulated within their particular framework, and each of them should give a certain level of theoretical success as well as shortcomings, such as the lack of an empirical confirmation.

In this chapter, we focus on the so-called \emph{nonlocal quantum gravity} (NLQG) \cite{Modesto:2011kw,Biswas:2011ar}.\footnote{Not to be confused with the homonymous proposal of \cite{Mashhoon:2017qyw,Mashhoon:2022ynk}. Unfortunately, there is no commonly established nomenclature classifying different nonlocal gravitational theories.} Although initially considered in the late 1980s, it has been a hot topic during the the last twenty years thanks to several indications that renormalizability and unitarity are indeed achieved. However, before we jump into the technical machinery of NLQG, we will shortly review gravity with higher-order derivatives to show what the main problems of local models are.

\subsection{Higher-derivative gravity} 

Higher-derivative theories constitute a generalization of GR in which one inserts extra curvature operators in the Einstein's action (\ref{EHAction}):
\begin{equation}
     S = \frac{1}{2\kappa^2}\int d^4x \sqrt{-g}\left[R + f(R,R_{\mu \nu}, R_{\mu \nu \rho \sigma})\right].
\end{equation}
In the attempt to develop a theory of quantum gravity, many types of models emerged, among which we highlight the following two. Although these theories improved our knowledge about quantum gravity, the presence of extra propagating degrees of freedom will become a problem both classically and at the quantum level.

\subsubsection{\texorpdfstring{$f(R)$}{f(R)} gravity}

$f(R)$ gravity is a class of theories characterized by a function of the Ricci scalar $R$. Proposed in the 1970s by Buchdahl \cite{Buchdahl:1983zz} and, in parallel, by Breizman, Gorovich and Sokolov \cite{Breizman1970POSSIBILITYOS}, these models have the action
\begin{equation}
      S = \frac{1}{2\kappa^2}\int d^4x \sqrt{-g} f(R) \: ,
\end{equation}
and one recovers Einstein's gravity when $f(R) = R$. A particular element of this class of theories has gained reputation in cosmology. Starobinsky inflation \cite{Starobinsky:1980te} is described by the Lagrangian density
\begin{equation}
    \mathcal{L} =  R + \frac{1}{6m^2}R^2,
\end{equation}
where $m$ is a mass scale. Curvature produces an accelerated expansion at early times that has important applications in primordial cosmology \cite{Vilenkin:1985md}.

\subsubsection{Stelle's gravity}

Stelle \cite{Stelle:1976gc} developed a theory of gravity with a quadratic action:
\begin{equation}
    S = \frac{1}{2\kappa^2}\int d^4x \sqrt{-g} \left( R + \alpha_1 R^2 + \alpha_2 R_{\mu \nu} R^{\mu \nu} + \alpha_3 R_{\mu \nu \rho \sigma}  R^{\mu \nu \rho \sigma} \right) \label{StelleAction}.
\end{equation}
This theory has interesting properties, for instance, that the additional coupling constants
\begin{equation*}
    \frac{\alpha_i}{2 \kappa^2}
\end{equation*} 
are dimensionless, whose implications in renormalization will be crucial as it will be explained in section \ref{sectionPowerCounting}.

Note that the Riemann-Riemann term can be removed in four dimensions since the Gauss--Bonnet action
\begin{equation}
    S = \frac{1}{2 \kappa^2}\int d^4x \sqrt{-g}\mathcal{L}_{\rm GB}\,,\qquad \mathcal{L}_{\rm GB} = R_{\mu \nu \rho \sigma}R^{\mu \nu \rho \sigma} - 4 R_{\mu \nu}R^{\mu \nu} + R^2 \label{eqGB} \:,
\end{equation}
is proportional to the Euler number $\chi({\cal M})$ that characterizes the topology of the manifold ${\cal M}$. The equations of motion derived from this action are trivial provided there is no topology change and no boundary terms. We will use this topological result to write the Riemann-Riemann term as a combination of the Ricci-Ricci scalar and the Ricci-Ricci tensor term.

\subsection{Unstable modes and Ostrogradski's theorem} \label{secOstro}

In 1850, Ostrogradski \cite{Ostrogradsky:1850fid} showed that higher-derivative classical theories have instabilities. This instability translates to a spontaneous decay of the vacuum, which results in the inability to consider such theories as physical ones to describe our universe. To see how this result arises, one can follow the Hamiltonian approach \cite{Woodard:2015zca} for two prototypical cases: theories with two derivatives and with four derivatives in the kinetic term.

\subsubsection{Two-derivative theories}

Considering a Lagrangian depending only on $x$ and $\Dot{x}$, one has that the Euler-Lagrange equations of this system are given by

\begin{equation}
   \frac{\partial L }{\partial x} - \frac{d}{dt}\frac{\partial L}{\partial \Dot{x}} = 0 \implies \Ddot{x} = F(x,\Dot{x}) \implies x(t) = x(t, x_0, \Dot{x}_0) \: ,
\end{equation}
and assuming non-degeneracy, i.e., $\partial^2L / \partial \Dot{x}^2 \neq 0$, one is able to write the Hamiltonian of the system through a Legendre transform. The Hamiltonian is conserved in the absence of explicit time dependence. Furthermore, one finds that this Hamiltonian is positive definite and therefore bounded from below.

In the context of field theory, we can reproduce these results using the simplest scalar field theory, i.e., through the Klein--Gordon equation
\begin{equation}
    S = - \int d^4x \left( \frac{1}{2}\partial^{\mu}\phi \partial_{\mu} \phi + \frac{1}{2}m^2 \phi^2  \right) \implies (\Box - m^2) \phi = 0 \: .
\end{equation}
For this case, one is able to construct the Hamiltonian density as
\begin{equation}
    \mathcal{H} = \frac{1}{2}\Pi^2 + \frac{1}{2} (\partial_i \phi \partial^i \phi) + \frac{1}{2}m^2 \phi^2 \geqslant 0 \label{eqHBounded} \: ,
\end{equation}
where $\Pi = \Dot{\phi}$ is the momentum associated with the field $\phi$. In (\ref{eqHBounded}), we see that $\mathcal{H}$ is bounded from below so that the theory is stable as will be explained shortly.

\subsubsection{Four-derivative theories}

Similarly, we can consider a Lagrangian depending only on $x$, $\Dot{x}$, and $\Ddot{x}$. In this case, one has that the Euler--Lagrange equations are

\begin{equation}
   \frac{\partial L }{\partial x} - \frac{d}{dt}\frac{\partial L}{\partial \Dot{x}} + \frac{d^2}{dt^2}\frac{\partial L}{\partial \Ddot{x}}  = 0 \implies \ddddot{x} = F(x,\dot{x}, \ddot{x}, \dddot{x}) \implies x(t) = x(t, x_0, \Dot{x}_0,  \ddot{x_0}, \dddot{x}_0),
\end{equation}
and assuming non-degeneracy, one is able to consistently write a Hamiltonian which is also constant in time if $L$ is time independent. However, the main difference here is that this new Hamiltonian is no longer bounded from below, as one can see analyzing the following four-derivative field theory:
\begin{equation}
    S = \int d^4x \left[ \frac{1}{2}\phi (\Box - \alpha \Box^2 ) \phi - \frac{1}{2}m^2 \phi^2  \right] \implies ( \Box - \alpha \Box^2 - m^2)\phi=0 \: . \label{eq2derivative}
\end{equation}
From this, one can build the associated Hamiltonian density through a Legendre transform to see that
\begin{equation}
    \mathcal{H} \propto \Pi_1 \dot{\phi} + \mathcal{O}[\Pi_2^2, (\nabla \phi)^2, \phi^2], \label{eqHUnBounded}
\end{equation}
where $\Pi_1 = \dot{\phi}-\alpha \dddot{\phi}$ and $\Pi_2 = \alpha \ddot{\phi}$. In this case, the linearity of $\mathcal{H}$ on $\phi$ and $\Pi_1$ prevents us to declare the Hamiltonian to be positive definite. Although this Hamiltonian is conserved, the linear term may take any value, even a negative one. Therefore, one concludes that $\mathcal{H}$ is unbounded from below.

In the massless case $m=0$ of (\ref{eq2derivative}), the bare propagator in momentum space is given by
\begin{equation}
    G(k^2) = - \frac{1}{k^2(1 + \alpha k^2)} = -\frac{1}{k^2}+ \frac{\alpha}{1+\alpha k^2} \label{eq2poles} \: .
\end{equation}
Identifying the poles of the propagator as the spectrum of the theory, we see that the presence of $\Box^2$ at the Lagrangian level introduces an extra pole in the propagator, leading to an additional massive mode of mass $m = 1/\sqrt{\alpha}$, provided $\alpha>0$. However, the different sign in the propagator indicates that this massive mode corresponds to a ghost \cite{Sbisa:2014pzo}. Furthermore, considering a homogeneous field $\phi (t)$, (\ref{eq2derivative}) becomes 
\begin{equation}
    \alpha \ddddot{\phi} - \ddot{\phi} = 0 \: ,
\end{equation}
whose real solution for $\alpha>0$  is
\begin{equation}
    \phi (t) = A \cosh{\frac{t}{\sqrt{\alpha}}} +  B \sinh{\frac{t}{\sqrt{\alpha}}} +C t +D \: ,
\end{equation}
where $A$, $B$, $C$ and $D$ are integration constants and this solution is clearly a non-oscillating and unbounded function.

The main consequence is that, already at the classical level, this system is unstable since the vacuum state can decay into excited states of particles and antiparticles contributing positively and negatively, respectively, to $\mathcal{H}$ \cite{Woodard:2015zca}. This decay is not only possible but favored from the entropy point of view \cite{Woodard:2015zca} and, therefore, one concludes that this kind of theory is irreconcilable with the observed universe because our ground state would be plagued by highly excited modes that do not decouple at some high energy, as it would happen in a stable theory. Moreover, this continuum decay is uncontrollable unlike to what happens, for instance, in QED. In addition, at the quantum level, one would obtain again a Hamiltonian that could be negative valued, being able to excite indefinitely a quantum state with unbounded energies and creating pairs of particle-antiparticle spontaneously that, in turn, would decay into higher-energy pairs of particles leading to a continuum decay of the vacuum state.

\subsubsection{Ghost modes}

These unstable additional modes are called \emph{ghost} modes since, as we have argued, the addition of higher-derivative terms in the Lagrangian comes at the price of producing extra degrees of freedom that are not observed in Nature. The kinetic term of these ghost fields appear in the Lagrangian with the wrong sign. At the quantum level, this translates into negative-norm states, leading to a violation of unitarity.

\section{Nonlocal classical scalar field theory}

Before we focus on nonlocal gravity, we study a nonlocal scalar field theory in order to illustrate how fundamental nonlocality manifests itself in the classical dynamics. We introduce the concepts of {form factors} and {kernel}, that will characterize the kind of nonlocality we have in our theory. We also show the problem of the initial conditions that these theories have to face and finally we expose how to solve it through the so-called \emph{diffusion method}.

\subsection{Motivation}

From the very beginning of the foundations of QFT, the main sectors of physical interest, such as QED, has been assumed to be \emph{local}, in the sense that the fields only depend on one spacetime coordinate $x^{\mu}$. This field theory, however, can be generalized to include nonlocality, in which some fields of the Lagrangian are evaluated at two different spacetime points. For instance, in the 1930s, Wataghin entertained the idea to introduce nonlocality to give a finite size to point-like particles \cite{Wataghin:1934ann}, and to do so he considered operators that were highly suppressed at large energies. For historical reasons, these operators were called \emph{form factors}, since their original purpose was to give `size' or `form' to dimensionless particles. 

Although local field theories prevailed, mainly because of the great predictivity of the Standard Model, nonlocal QFT has been an area of research in the second half of the XX century and many authors have applied these ideas to quantum scalar fields, gauge fields, gravity and cosmology.

In general, one distinguishes between two kinds of nonlocality: the nonlocality induced by a Lagrangian valued at a field that depends on different points on spacetime $\mathcal{L}[\phi(x,y)]$, or the nonlocality arising in a Lagrangian that depends on multiple fields evaluated at different spacetime points, i.e., $\mathcal{L}[\phi(x),\phi(y)]$. Whereas the former nonlocality appears in the formulation of standard QFT, e.g., in multi-point multi-tensor objects such as Green's functions, here we focus on the latter, which is present in many areas of theoretical physics such as noncommutative QFT \cite{Snyder:1946qz}, string field theory \cite{Eliezer:1989cr}, effective field theories \cite{Barvinsky:1985an}, and conformal field theory \cite{Wess:1971yu,Witten:1983tw,Novikov}.

\subsection{Nonlocality}\label{noloca}

In the most common local field theory, the action and the equations of motion of the system are given by
\begin{equation}
    S = \int d^4x \left[ \frac{1}{2}\phi(x) \Box \phi(x) - V(\phi)\right] \quad \implies \quad \Box \phi (x) - \frac{dV}{d\phi(x)} = 0   \: . \label{LocalScalarAction}
\end{equation}
One may introduce fundamental nonlocality by considering a more general function $\Box \to \gamma(\Box)$. More precisely, we consider not a finite sum of higher-derivative terms, but an infinite sum of these operators in the following form:
\begin{equation}
    \gamma (\Box) = \sum_{n=0}^{\infty} c_n \Box^n \label{seriesRepFF},
\end{equation}
so that the previous action (\ref{LocalScalarAction}) becomes
\begin{equation}
    S = \int d^4x \left[ \frac{1}{2}\phi(x) \gamma(\Box) \phi(x) - V(\phi)\right] \: .
\end{equation}
The equations of motion of this theory can be obtained via
\begin{equation}
    \frac{\delta S}{\delta \phi(x)} = 0 \quad \implies \quad \gamma ( \Box) \phi (x) - \frac{d V}{d \phi (x)} = 0 \: .
\end{equation}
The connection of this class of theories with nonlocality, although at first sight not apparent, can be seen through the following algebraic manipulation:
\begin{equation}
\begin{split}
     \gamma (\Box) \phi (x) &=  \int d^4 k \gamma (-k^2) \delta^{(4)}(k^{\mu}-i \nabla^{\mu}) \phi (x) \\
     & =  \int d^4k \int d^4y e^{-i y \cdot k} K(y) \delta^{(4)}(k^{\mu}-i \nabla^{\mu}) \phi (x)\\
     & = \int d^4y K(y) e^{y \cdot \nabla}\phi (x)\\
     & = \int d^4y K(y) \phi (x+y) \\
     & = \int d^4y K(y-x) \phi (y)  \: ,\\ \label{IntegralRepresentationFF}
\end{split}
\end{equation}
where in the last equality we have applied a change of variables and we have used the Fourier transform along the way. We see that the action $\gamma (\Box)$ on $\phi (x)$ applies a \emph{delocalization} of the field, so that the kinetic term 
\begin{equation}
    \phi (x) \gamma (\Box) \phi (x) = \int d^4y \phi (x) K (y-x) \phi (y) \label{eqKernel} \: ,
\end{equation}
relates two scalar fields evaluated at two different spacetime points via the delocalization kernel $K(y-x)$. Thus, and somewhat surprisingly, any kinetic term can be formally regarded as a nonlocal interaction. We call $\gamma (\Box)$ \emph{form factor} and this is precisely what Wataghin was interested in. From (\ref{eqKernel}), one sees that
\begin{equation}
    K(x) = \gamma (\Box_x) \delta^{(4)}(x) = \int \frac{d^4k}{(2 \pi)^4}e^{-ik \cdot x } \gamma ( - k^2) \label{MinkFTFF} \: .
\end{equation}
The form of $\gamma (\Box)$ determines the particle spectrum of the free theory via the dispersion relation
\begin{equation}
    \gamma(-k^2) = 0 \:.
\end{equation}
For instance, one could consider the usual local scalar field theory (\ref{LocalScalarAction}) with $V=0$, and in this case $\gamma (-k^2) = - k^2 = 0$, that is the dispersion relation for the massless scalar field theory. The integral representation of the form factor also works even for local higher-derivative theories. Then, the result is special and the kernel $K$ is just a sum of Dirac deltas and their derivatives of finite order. In general, for nonlocal theories $K$ is a smooth continuous function of the difference of the coordinates $x-y$ (because of translational invariance). 

In general, from a kernel $K(x,y)$ defined at two different spacetime points and composing the operator $\int dy\phi(x)K(x,y)\phi(y)$, it is possible to obtain a kinetic term $\phi(x)\gamma(\Box)\phi(x)$ only in very special situations. From this observation, another classification of nonlocality distinguishes three types.
\begin{itemize}
\item \emph{Weak nonlocality}, when the kinetic term is an analytic function of the Laplace--Beltrami operator. Then, the form factor $\gamma(z)$ admits a regular Taylor series expansion around the infrared (IR) point $\Box=z=0$. 
\item \emph{Strong nonlocality}, with a form factor of the type $\Box^{-1}$ or other inverse powers $\Box^{-n}$ singular at $z=\Box=0$.
\item \emph{Very strong nonlocality}, when the kinetic term is made of an integral kernel $K(x,y)$ not convertible into a derivative operator $\gamma(\Box)$. (In this case, the function $\gamma(\Box)$ can be written only formally and it really does not exist due to convergence problems of the inverse Fourier transform from (\ref{MinkFTFF})). 
\end{itemize}
Inside the class of weakly nonlocal form factors, we can distinguish those which are exponential in powers of the $\Box$ \cite{Wataghin:1934ann,Krasnikov:1987yj,Eliezer:1989cr}, those which are asymptotically polynomial in the ultraviolet (UV) regime along the real axis \cite{Kuzmin:1989sp,Tomboulis:1997gg,Modesto:2011kw}, and also fractional powers of the $\Box$ \cite{Bollini:1964,Bollini:1991fp,Calcagni:2021ljs}. NLQG is characterized by weak nonlocality and asymptotically polynomial form factors \cite{Kuzmin:1989sp,Tomboulis:1997gg,Modesto:2011kw}, although one can formulate quantum gravity also with fractional operators \cite{Calcagni:2021aap,Calcagni:2022shb}. Higher-derivative local theories are when the form factor is a finite polynomial and its exponents are positive integer numbers \cite{Asorey:1996hz,Accioly:2002tz}.

\subsection{Representations of form factors}
\label{SectionAppendix}

In general, one may define a nonlocal operator in two different ways or \emph{representations}: the \emph{integral representation} or the \emph{series representation}. It is expected that weakly nonlocal form factors admit both representations, although they can give very different results when applied to certain seed functions. As an example, apply the cosmological $\Box$ operator with Hubble parameter $H\propto 1/t$ on a power law $t^n$: The series representation of $e^\Box$ fails to converge, while the integral one gives a finite answer \cite{Calcagni:2007ru}. 

\subsubsection{Series representation}

Already introduced in eq.\ (\ref{seriesRepFF}), this is the most common representation and it is based on the decomposition of the nonlocal form factor $\gamma (\Box)$ as an infinite sum of the Laplace--Beltrami operator. Analogously to what happens when trying to Taylor expand a function around a value which is not contained in the domain of the function, this representation is only valid for form factors $\gamma(z)$ that are regular at $z=0$. For instance, the form factor $\Box^{-1}$ does not admit a series representation without regularization. Finally, we point out that this representation has the same form and structure in any spacetime background.

\subsubsection{Integral representation}

The integral representation has been introduced in (\ref{IntegralRepresentationFF}) in the scalar case using Minkowski spacetime as the background of the theory, with the kernel function, which is defined in (\ref{MinkFTFF}). This representation is only valid in flat spacetime in Cartesian coordinates, since in general the Laplace--Beltrami operator is accompanied by terms dependent on the connection of the manifold. In this sense, we say that the decomposition in terms of the exponentials $\exp(\pm i k \cdot x)$ is no longer useful for such representation since they are no longer eigenfunctions of $\Box$. 

Therefore, when constructing explicit solutions of the nonlocal gravitational theory, first of all, one needs to build the integral representation of the nonlocal form factor out of the two eigenstates $\mathcal{B}_l$ of the Laplace--Beltrami operator in curved spacetime following this recipe \cite{Calcagni:2007ru}:
\begin{enumerate}
    \item Find the two eigenstates of the Laplace--Beltrami operator by solving the second-order differential equation
    \begin{equation}
        (\Box + k^2)\mathcal{B}_l(k,x) = 0, \quad \quad l=1,2 \: ,
    \end{equation}
    where $k$ is either real or purely imaginary.
    \item Write the field as a linear superposition of the eigenstates $\mathcal{B}_l$. For the scalar case we may define two integral transformations of the field as
    \begin{equation}
        \bar \phi_l(k) = \int d^4x \mathcal{B}_l(k,x)g_l(x)\phi(x) \: ,
    \end{equation}
    with $g_l$ being certain weights. Thus, the scalar field may be written as
    \begin{equation}
        \phi (x) = \int d^4k \left[ c_1\mathcal{B}_1(k,x)\bar{\phi}_1(k) + c_2\mathcal{B}_2(k,x)\bar{\phi}_2(k)  \right] \: ,
    \end{equation}
    where $c_{1,2}\in \mathbb{C}$ are constants and different for each particular field.
				\item Write the nonlocal form factor acting on the scalar field as
    \begin{equation}
        \gamma (\Box) \phi (x) = c_1 \bar{\phi}_1(x) + c_2 \bar{\phi}_2(x) \: ,
    \end{equation}
    where
    \begin{equation}
        \bar{\phi}_l(x) = \int d^4k \mathcal{B}_l(k,x)\gamma(-k^2)\bar{\phi}_l(k) \: .
    \end{equation}
\end{enumerate}

\subsection{Form factors} \label{SecFF}

We are interested in making our quantum theory renormalizable but, as one could imagine, not all form factors will be viable for this purpose. In this section, we introduce form factors of interest in quantum gravity, and since the energy dimension of $\Box$ is $[\Box]=2$, we will have to introduce a characteristic length in the theory, denoted by $l_*$. We will make explicit the momentum dependence of these form factors in the UV and how this behaviour for large momenta affects the renormalization of the theory.

\subsubsection{Exponential form factor}

This form factor can be written as
\begin{equation}
    \gamma ( \Box) = ( \Box - m^2) e^{-l_*^2 \Box},
\end{equation}
and the bare propagator (the inverse of $\gamma$) in momentum space is given by
\begin{equation}
    G(k^2) = - \frac{e^{-l_*^2 k^2}}{k^2+m^2} \: .
    \label{ff26}
\end{equation}
This expression is enlightening because one sees that it has only one real pole at $k^2 = - m^2$, so that the exponential form factor, unlike what happens with higher-derivative terms  in (\ref{eq2poles}), does not introduce additional modes into the spectrum of the theory.

One can also study the action of exponential form factors on some well-behaved functions that tend to zero very fast when their argument is sent to infinity (UV limit). One such example of the probe function is, in the one-dimensional case, the Gaussian function $f(x)=\exp\left[A (x-x_0)^2 +B\right]$, where $A<0$, $B$, and $x_0$ are constants; this example can be easily generalized to the higher-dimensional case. The result of the action of the spacetime operator $\exp(a\partial_x^2+b \partial_x +c)$ on $f$ is, provided $1-4aA>0$,
\begin{equation}
 \exp(a\partial_x^2+b \partial_x +c) f(x) =  \frac{\exp \left\{-\frac{A \left[-4 a (B+c)+b^2-2 b
   (x-x_0)+(x-x_0)^2\right]+B+c}{4 a A-1}\right\}}{\sqrt{1-4 a A}},
\end{equation}
which is still a Gaussian function suppressed at spatial infinity. We see that the action of the exponential form factor translates into the transformation of the coefficients characterizing the original Gaussian:
\begin{equation}
 A'=\frac{A}{4 a A-1},\qquad x_0'=x_0+b,\qquad B'=-\frac{1}{2} \ln (1-4 a A)+B+c\,.
\end{equation}
Hence one can conclude that these are simple operations on the parabola defining the Gaussian, like shifting its vertex horizontally (with the $b$ parameter), vertically (with the $c$ parameter), and changing the shape or opening thereof (with the $a$ parameter). 

Therefore, by this example one unambiguously understands the action of nonlocal exponential form factors of basic one-dimensional differential operators on highly suppressed probe functions as just changing the shape and positions of these functions without modifying their asymptotic fall-off properties. For a general function with different infinite asymptotics, the action of the form factor may not be so well defined and may possess some ambiguities.

\subsubsection{More general form factors}

In NLQG we are interested mainly in four nonlocal operators, called Wataghin, Krasnikov, Kuz'min, and Tomboulis form factors. Here we write their definitions and their properties in the UV. We forewarn the reader that, while all these form factors work well for a scalar theory, only the last two (asymptotically polynomial) are eventually under full control in quantum gravity, since in the presence of gauge or diffeomorphism invariance there are residual divergence in loop diagrams when using the first two form factors (exponential). All of them can can be written as
\begin{equation}
    \gamma (\Box) = \frac{e^{\text{H}(\Box)}-1}{\Box} \label{APFF} \: ,
\end{equation}
where $\frac{1}{\Box}$ stands for the inverse Laplace--Beltrami operator (always absorbed by the leading $O(\Box)$ term in the numerator; hence there is no strong nonlocality here) and the function $\text{H}(\Box)$ will be different for each case. Equation (\ref{APFF}) has the same purpose as in (\ref{ff26}) not to introduce new poles, except the already existing one of the massless spin-2 graviton from the Einstein--Hilbert term in the case of quantum gravity.\footnote{If this assumption is not used, then one can even fancy models without any physical degree of freedom where the kinetic has the simple form $\exp \text{H}(\Box)$. $p$-adic models have this structure \cite{Brekke:1987ptq}.} 

\subsubsection{Wataghin/minimal form factor}

Characterized by $\text{H}_{\text{Wat}}(\Box) = -l_*^2 \Box$, it is the form factor that has been used the most due its simplicity, although nowadays the emphasis in NLQG has been displaced to asymptotically polynomial form factors. Its expression is given by
\begin{equation}
    \gamma (\Box) = \frac{e^{-l_*^2\Box}-1}{\Box} \label{WataghinFF} \: .
\end{equation}

\subsubsection{Krasnikov form factor}

Defined in 1987 by Krasnikov \cite{Krasnikov:1987yj}, it is characterized by $\text{H}_{\text{Kras}}(\Box) = l_*^4 \Box^2$:
\begin{equation}
    \gamma (\Box) = \frac{e^{l_*^4\Box^2}-1}{\Box} \label{KrasnikovFF} \: .
\end{equation}

\subsubsection{Kuz'min form factor}

Now we introduce a class of asymptotically polynomial form factors whose asymptotic behaviour is very different from the previously considered exponential type. We begin with form factors being asymptotically monomial, where the characteristic monomial is given by $m(z)=a z^{n_{\rm deg}}$ with $a$ a constant and $n_{\rm deg}$ the degree of the monomial given by an integer number.

The function $\text{H}(z)$ in these asymptotically monomial form factors must satisfy certain conditions.
\begin{description}
    \item[1.] It is real and positive along the real axis.
    \item[2.] $\text{H}(0) = 0$ such that $e^{\text{H}(0)}=1$.
    \item[3.] It grows no faster than a monomial of finite degree $n_{\rm deg}$ when one takes $|z| \to +\infty$ in a particular conical region $\mathcal{C}$ of the complex plane. This region is characterized by an angle $\theta$ with respect to the real axis. Thus we can write this condition mathematically as
    \begin{equation}
        e^{\text{H}(z)} \stackrel{|z| \to \infty}{\simeq} |z|^{n_{\rm deg}} \quad \quad z \in \mathcal{C}, \quad -\theta < \text{arg}(z) < \theta \: .
    \end{equation}
    \item[4.] In the conical region
    \begin{equation}\lim_{|z| \to \infty}
        \frac{e^{\text{H}(z)}- |z|^{n_{\rm deg}}}{|z|^{n_{\rm deg}}} z^n  = 0 \quad \quad \forall n \in \mathbb{N} \: .
    \end{equation}
\end{description}
The rationale behind asymptotically monomial form factors will be clearer in section \ref{SecRenormalization}, where the UV behaviour, dictated by the order $n_{\rm deg}$ of the polynomial, will be essential to achieve renormalizability.

Kuz'min form factor is given by formula (\ref{APFF}) with
\begin{equation}
    \text{H}(\Box) = \alpha [\ln(m(\Box))+ \Gamma[0,m(\Box)]+\gamma_{\rm E}] \equiv \text{H}_{\text{Kuz}}(\Box) \: ,
\end{equation}
where $\gamma_{\rm E}$ is the Euler--Mascheroni constant, $m(\Box)$ is a real monomial ($a\in\mathbb{R}$) of degree $n_{\rm deg} \geqslant 1$, $\alpha$ is a real constant such that $\alpha \geqslant 3$ (in order to have one-loop super-renormalizability, i.e., perturbative UV divergences only at the one-loop level) and $\Gamma$ is the upper incomplete gamma function defined as
\begin{equation}
    \Gamma(0,z) = \int_z^{\infty}dx \frac{e^{-x}}{x} \label{IncompleteGamma} \: .
\end{equation}
More precisely, Kuz'min \cite{Kuzmin:1989sp} introduced this form factor in 1989 with $m(\Box) = - l_*^2 \Box$, hence $n_{\rm deg}=1$. This construction can be generalized to a polynomial asymptotic behaviour, as explained in section \ref{generalpol} below.

\subsubsection{Tomboulis form factor}

\begin{figure}[H]
    \centering
    \includegraphics[width = 0.7\textwidth]{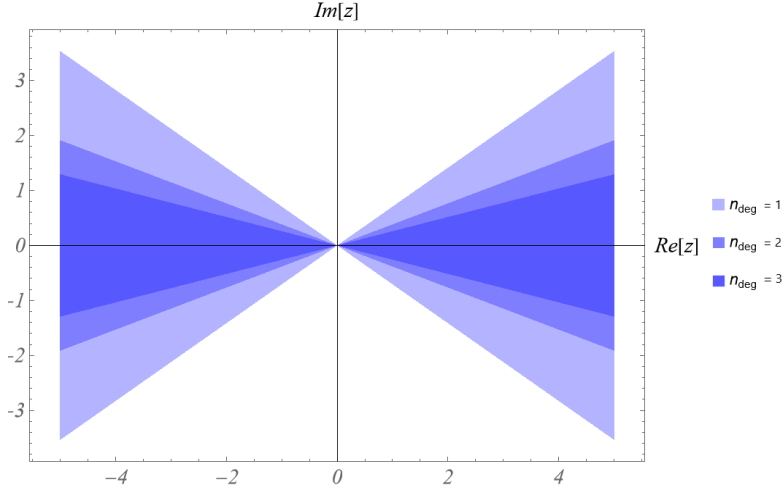}
    \caption{Conical regions $\mathcal{C}$ of the Tomboulis form factor $\text{H}_{\text{Tom}}$ for $n_{\rm deg}=1,2,3$.}
    \label{fig:ConicalRegion}
\end{figure}

This form factor is defined with the asymptotic polynomial $p=p(z)$ by
\begin{equation}
    \text{H}(\Box) = \frac{1}{2}\{\ln[p^2(\Box)] + \Gamma[0,p^2(\Box)] + \gamma_{\rm E}\} \equiv \text{H}_{\text{Tom}}(\Box) \: ,
\end{equation}
which is defined in the conical regions
\begin{equation}
    \mathcal{C} = \{ z | \quad - \theta < \text{arg}(z) < \theta \quad  \cup \quad \pi - \theta < \text{arg}(z) < \pi + \theta \}, \quad \quad \theta = \frac{\pi}{4 n_{\rm deg}} \: ,
\end{equation}
depicted in figure \ref{fig:ConicalRegion} for the cases of monomials with degrees $n_{\rm deg} = 1, 2, 3$ respectively.

Asymptotically polynomial form factors are not everywhere asymptotic to a polynomial in the UV regime $|z|\gg1$. They are such only inside the conical regions along the real axis. Outside these regions on the complex plane, they may have other higher-order asymptotics such as $\exp(\exp z)$, for example. Moreover, all these non-trivial entire functions have an essential singularity at the complex infinity $z=\infty_c$. At any rate, form factors like Kuz'min and Tomboulis can be defined as entire functions also outside these conical regions and outside the essential singularity at infinity.

\subsubsection{General asymptotically polynomial form factor}
\label{generalpol}

We view an asymptotically polynomial form factor as a general analytic complex function of Fourier space momentum $k_{\mu}$. Covariant form factors depend on $\Box=\to-k^{2}=-g^{\mu\nu}k_{\mu}k_{\nu}$: ${\gamma}={\gamma}(k^{2})$. In Euclidean domain, $k^{2}=\delta^{\mu\nu}k_{\mu}k_{\nu}$ and, calling $z=k^{2}$, we have
    \begin{equation}
{\gamma}={\gamma}(z)\,.
    \end{equation}
For $z\gg1$ and in the asymptotic limits of $|z|\to+\infty$ in the conical region $\cal C$ along the real positive axis, we should have asymptotics to a polynomial $p=p(z)$ with a degree $n_{\rm deg}\in\mathbb{N}$:
\begin{equation}
p(z)=a_{n_{\rm deg}}z^{n_{\rm deg}}+\sum_{i=0}^{n_{\rm deg}-1}a_{i}z^{i},
\end{equation}
where $a_{n_{\rm deg}}\neq0$. Then, in the UV we have a higher-derivative local theory.
		
In Euclidean signature, the UV regime is only when $k^{2}\gg1$ and this regime is responsible for the UV properties of the theory such as the structure of UV divergences, renormalizability, super-renormalizability or UV-finiteness. In Lorentzian signature, the situation is more delicate. We have $k^{2}=g^{\mu\nu}k_{\mu}k_{\nu}$ with metric signature $(-,+,+,+)$. The deep (physical) UV regime is then defined as two regimes, $k^{2}\to+\infty$ ($|k^{2}|\gg1$ on the real line) and $k^{2}\to-\infty$ ($|k^{2}|\gg1$ on the real line and negative value for the Lorentzian square $k^{2}$). However, a possible UV regime could also arise when $|\bm{k}|=|k^{0}|\to+\infty$, i.e., when we are on the light cone $k^{2}=-\left(k^{0}\right)^{2}+|\bm{k}|^{2}=0$. In such a special condition, the analysis of the UV behaviour of the form factor is carried out at the argument $z=k^{2}=0$. The UV divergences which could arise in such situation would not be preserving Lorentz symmetry, since the condition on the component $k^{0}\gg1$ is not imposed on the full Lorentz-invariant length of the four-vector $k^\mu$.

For the sake of good renormalizability properties of the Lorentzian theory, one should require that the asymptotically polynomial
behaviour holds in the two regimes $k^{2}\to\pm\infty$ with the same polynomial and that the conical regions are symmetric with respect to the origin point. The asymptotics is such that, for $|z|\to+\infty$, we should have in the conical regions that ${\gamma}(z)\simeq p(z)$:
\begin{equation}
\lim_{\stackrel{|z|\to+\infty}{z\in\mathcal{C}}}\frac{{\gamma}(z)}{p(z)}=1\,.
\end{equation}
The difference should be suppressed as
    \begin{equation}
\lim_{\stackrel{|z|\to+\infty}{z\in\mathcal{C}}}z^{\alpha}\left[{\gamma}(z)-p(z)\right]=0\,,
    \end{equation}
for any $\alpha\in\mathbb{R}$, especially for $\alpha>0$. Then, it is guaranteed that the analysis of perturbative UV divergences
gives the same result in both regimes $k^{2}\to\pm\infty$ and coincides with the analysis of infinities in local higher-derivative theory described by the polynomial $p(z)$.

A more general condition for the natural order $n$ of the form factor
understood as the complex function is
    \begin{equation}
\lim_{\stackrel{|z|\to+\infty}{z\in\mathcal{C}}}\frac{\ln{\gamma}}{\ln z}=n\,,
    \end{equation}
so that the function here has the order $n$ in the analysis of the
asymptotics of complex entire functions. In this case, only the leading UV divergences
could be captured correctly by a local higher-derivative theory with the
monomial $a_{n}z^{n}$. However, even this general case and definition does not imply that
    \begin{equation}
\lim_{\stackrel{|z|\to+\infty}{z\in\mathcal{C}}}\frac{{\gamma}}{a_{n}z^{n}}=1.
    \end{equation}
    The last equality holds for truly asymptotically polynomial form factors with degree $n_{\rm deg} = n$, while for general complex functions of order $n$ it may be not satisfied.

\subsubsection{Summary of form factors} \label{subsectionSummary}

We can summarize the UV properties of the previous form factors.
\begin{enumerate}
    \item Wataghin, Krasnikov, Kuz'min, and Tomboulis form factors diverge in the UV in Euclidean momentum space, i.e., $k_E^0 = ik^0$, as one can see in figure \ref{fig:Euclidean}.
\begin{figure}[H]
    \centering
    \includegraphics[width = 0.7\textwidth]{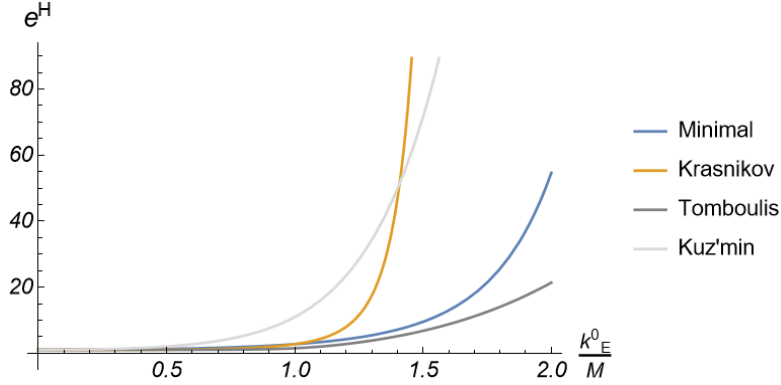}
    \caption{ $\text{exp}(\text{H})$ as a function of the Euclidean momentum $k^0_E$, where we have defined the energy scale $M = 1/l_*$ and taken $\alpha = 3$ and $m(\square)=-\square$ for Kuz'min form factor and a monomial with $n_{\rm deg} = 2$ for Tomboulis form factor.}
    \label{fig:Euclidean}
\end{figure}
Therefore, since all of them blow up in the UV, the associated propagator, which is the inverse of the form factor, will be highly suppressed in the UV, thus facilitating the convergence of Feynman diagrams for the scalar case. (The case of gauge theories and gravitation is more complicated since also interaction vertices include form factors.) Furthermore, the huge growth of the form factors in Euclidean momentum space implies asymptotic freedom: since the kinetic term dominates in the UV when considering its running along the renormalization-group energy scale, interactions become negligible and the resulting theory is asymptotically free. We will come back to this point in section \ref{secAF}. We also emphasize that these form factors are not UV-divergent on the light cone where the massless on-shell dispersion relation$k^2=-(k^{0})^{2}+|\bm{k}|^{2}=0$ is satisfied, since $\text{H}(0)=0$. Hence there is a suppression of the propagator in the UV everywhere except on the light cone. 
    \item Their behaviour is different in Lorentzian momentum space. In particular we have that
    \begin{equation}
        \lim_{k^0 \to \infty} \text{H}(- k^2) = - \infty \quad \quad \text{for Wataghin and Kuz'min form factors,}
    \end{equation}
    and
    \begin{equation}
        \lim_{k^0 \to \infty} \text{H}(- k^2) = + \infty \quad \quad \text{for Krasnikov and Tomboulis form factors,}
        \label{eq45}
    \end{equation}
    as one sees in figure \ref{fig:Lorentzian}. This difference occurs because the first group of form factors is sensitive to the sign of their argument $z=k^2$, while in the second group the dependence of the argument is always quadratic ($z^2$ or $p^2(z)$) and the limits $k^2\to\pm\infty$ give the same result.
\begin{figure}[H]
    \centering
    \includegraphics[width = 0.7\textwidth]{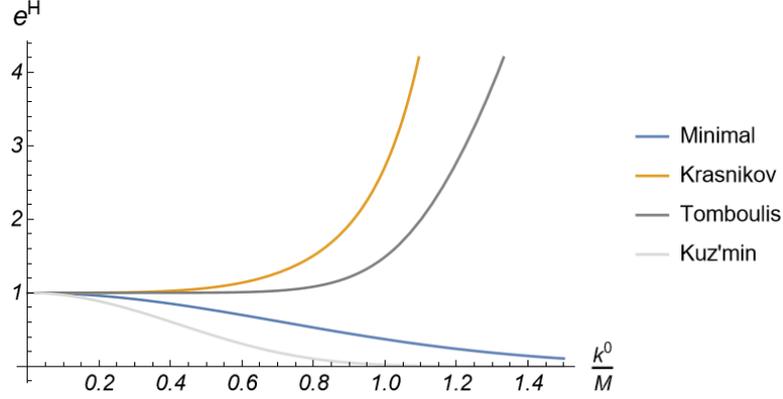}
    \caption{$\text{exp}(\text{H})$ as a function of the Lorentzian momentum $k^0$, where we have defined the energy scale $M = 1/l_*$ and taken $\alpha = 3$ and $m(\square)=-\square$ for Kuz'min form factor and a monomial with $n_{\rm deg} = 2$ for Tomboulis form factor.}
    \label{fig:Lorentzian}
\end{figure}
\end{enumerate}

\subsection{Stability and initial conditions}

We saw in section \ref{secOstro} that an immediate consequence of adding higher-derivative terms to Einstein's theory (\ref{EHAction}) is the intrusion of extra degrees of freedom, some of which in the form of ghost fields. Through the use of the nonlocal form factors introduced in section \ref{SecFF}, we commented on the possible amelioration of the divergences of the Feynman diagrams. However, since all the form factors that we consider can be expanded in a power series of the Laplace--Beltrami operator, we might wonder how this procedure ensures that they do not actually add an infinite number of extra propagating modes. Moreover, the absence of ghosts is clear from the absence of extra poles but, looking from the side of the infinite sum of powers of $\Box$, how is it possible that these manage to cancel any Ostrogradski instability? 

In any classical theory with $n$ derivatives in the Lagrangian, one needs to specify $2n$ initial conditions to solve uniquely the physical system. However, in the nonlocal case, the kinetic term contains an infinite number of derivatives, so that one might say that we need an input of an infinite number of initial conditions at $t=t_0$:
\begin{equation}
    \phi^{(n)}(t_0,\bm{x}) \quad \quad \forall n \in \mathbb{N} \: . \label{InfiniteIC}
\end{equation}
However, these initial conditions are precisely what we need to construct the solution $\phi(t,\bm{x})$ as a power series, provided that the solution is real and analytic (which is different from the requirement that the solution be smooth and that it smoothly depend on initial data):
\con
\begin{equation}
    \phi (t,\bm{x}) = \sum_{n=0}^{\infty}\frac{ \phi^{(n)}(t_0,\bm{x})}{n!}(t-t_0)^n \: . \label{eqPowerSeries}
\end{equation}
Thus, we face a paradox \cite{Moeller:2002vx}. We need an infinite number of initial conditions to specify the solution $\phi (t,\bm{x})$ but, once we have them, we can directly construct our field via the expression (\ref{eqPowerSeries}) assuming reality and analyticity at $t=t_0$. To fully solve the problem of initial conditions, we should already know the solution!

\subsection{Diffusion method} \label{sectionDEM}

We can overcome this issue using the diffusion method, initially used in the context of string field theory \cite{Calcagni:2007ru,Calcagni:2009jb}. For simplicity, we illustrate the method for Wataghin form factor \cite{Calcagni:2018lyd} but, after adaptations, it holds also for asymptotically polynomial form factors \cite{Calcagni:2018gke}. In essence, one promotes our four-dimensional field $\phi(t,\bm{x})$ into a field $\Phi (r,t,\bm{x})$ depending on a fictitious extra dimension $r$, with the constraint that $\Phi (r,t,\bm{x})$ obeys the diffusion equation on the $r$ coordinate:
\begin{equation}\label{difeq}
    (\partial_r - \Box)\Phi(r,t,\bm{x}) = 0 \: .
\end{equation}
Together with this extra dimension, one introduces an auxiliary field $\chi (r,t,\bm{x})$ to impose the diffusion equation on $\Phi (r,t,\bm{x})$ that, in turn, is constrained by $\chi = \Box \Phi$. In the above equation and everywhere below, $\Box$ is understood as the four-dimensional operator. Also, one can write an action with measure $d^4x dr$ \cite{Calcagni:2018lyd} but it is paramount to stress that the $r$-dependent system is never conceived as a physical five-dimensional extension of the original nonlocal theory. The extra direction is flat (it does not come with any warp factor) and there is no attempt to implement five-dimensional covariance, nor to study the localized system as a five-dimensional QFT to be made unitary or renormalizable. Rather, the four-dimensional system is regarded as living on a flat slice at some special value $r=\tilde\beta r_*$ in this abstract ambient space, where $\tilde\beta > 0$. 

The motivation to constrain $\Phi(r,t,\bm{x})$ via a diffusion equation is that we can treat the exponential form factor as a translation in the extra dimension $r$:
\begin{equation}
    e^{-l_*^2 \Box}\Phi(r,t,\bm{x}) = e^{- l_*^2 \partial_r}\Phi(r,t,\bm{x}) = \Phi(r- l_*^2,t,\bm{x}) \: ,
\end{equation}
and we recover the physical field configuration in four dimensions $\phi(t,\bm{x})$ for a specific value of the extra component $r$, i.e., at the particular slice location in $r$ proportional to $r_* = l_*^2$. Note that $[r]=-2$, so that strictly speaking it is not on the same level as a coordinate. This choice is dictated to make the diffusion equation (\ref{difeq}) parameter free, with diffusion coefficient equal to 1.


\subsubsection{Procedure}

To illustrate how this method works, we focus on the particular nonlocal system
\begin{equation}
    S_{\phi} = \int d^4x \mathcal{L}, \quad \quad \mathcal{L} = \frac{1}{2}\phi \Box e^{-r_*\Box}\phi - V[\phi] \label{nonlocalsys} \: ,
\end{equation}
whose equation of motion is
\begin{equation}
    \Box e^{-r_*\Box}\phi - V'[\phi] = 0\label{nonlocaleom} \: .
\end{equation}
In the definition of this model, we assume that the potential is a local function of the field $\phi=\phi(x)=\phi(t,\bm{x})$, without derivatives.

First of all, we introduce two fields $\Phi (r,t,\bm{x})$ and $\chi (r,t,\bm{x})$ local in four-dimensional spacetime directions (i.e., their four-dimensional dynamics in $x$ is local), while the nonlocality is completely transferred to the unphysical extra dimension $r$. By definition, we recover the physical dynamics of the nonlocal system when evaluating the field $\Phi$ at the special slice $r=\tilde\beta r_*$ for a certain $\tilde\beta > 0$, such that $\Phi (\tilde\beta r_*,t,\bm{x}) = \phi (t,\bm{x})$. Because of the locality of the dynamics of the field $\Phi (r,t,\bm{x})$ in the spacetime coordinates, we only require a finite number of initial conditions for the field $\phi (t,\bm{x})$.

One can build a suitable Lagrangian for the localized system \cite{Calcagni:2018lyd}
\begin{equation}
    S[\chi,\Phi] = \int d^4xdr (\mathcal{L}_{\chi} + \mathcal{L}_{\Phi}) \label{LocalizedAction5D} \: ,
\end{equation}
where
\begin{equation}
    \mathcal{L}_{\Phi} = \frac{1}{2}\Phi (r,x) \Box \Phi (r-r_*,x) - V[\Phi(r,x)] \: ,
\end{equation}
\begin{equation}
    \mathcal{L}_{\chi} = \frac{1}{2}\int_0^{r_*}dq \chi(r-q,x)(\partial_{q}-\Box)\Phi(r+q-r_*,x) \label{ScalarLagrangianChi} \: ,
\end{equation}
and we have encapsulated the time component $t$ and the spatial coordinates $\bm{x}$ into the label $x$. Notice that this action entails a local dynamics in the four-dimensional spacetime coordinates but nonlocal in the unphysical coordinate $r$. The equations of motion of this action are given by
\begin{equation}
    \frac{\delta S}{\delta \chi} = 0\,, \qquad \frac{\delta S}{\delta \Phi} = 0 \: ,
\end{equation}
leading to the expressions \cite{Calcagni:2018lyd}
\begin{align}
     0 \:= \:&\: (\partial_r - \Box) \Phi(r,x)  \label{eqDifPhi} \: , \\
     0 \:= \:&\: (\partial_r - \Box) \chi(r,x) \: ,  \\
     0 \:= \:&\: \frac{1}{2}[\Box \Phi(r-r_*,x) + \chi(r-r_*,x)] + \frac{1}{2}[\Box \Phi (r+r_*,x) - \chi(r+r_*,x)] - V'[\Phi(r,x)] \: ,  \label{eqCHI}
\end{align}
where $V'[\Phi(r,x)] = dV/d\Phi(r,x)$.

Both fields $\Phi$ and $\chi$ follow a diffusion equation. The solutions to these equations of motion are not uniquely determined. However, we have a freedom left to impose an extra constraint
\begin{equation}
    \Box \Phi(r,x) = \chi(r,x) \: ,\label{eqConstraint}
\end{equation} 
for any value of $r$, so that the auxiliary field $\chi$ freezes out at the physical slice and equation (\ref{eqCHI}) becomes
\begin{equation}
\begin{split}
    0\: =\: & \Box \Phi (r-r_*,x) - V'[\Phi(r,x)] = \Box e^{- r_* \partial_r} \Phi(r,x) -  V'[\Phi(r,x)] \\\: =\:&  \Box e^{-r_*\Box} \Phi(r,x) -  V'[\Phi(r,x)]\: .
\end{split}
\end{equation}
Evaluating this expression at $r = \tilde\beta r_*$, i.e., the  slice where $ \Phi ( \tilde\beta r_*,x) = \phi (x)$, one recovers the equation of motion of the physical nonlocal system (\ref{nonlocaleom}). 

Although the equivalence of both systems has been established for a particular theory (\ref{nonlocalsys}) with an exponential form factor, this result can be generalized for other theories such as those with an exponential-polynomial form factor \cite{Calcagni:2018lyd} or an asymptotically polynomial form factor \cite{Calcagni:2018gke}.

\subsubsection{Initial conditions, degrees of freedom and absence of ghosts}

Given a nonlocal system, we can always write down a localized system whose solutions to the equations of motion at the physical slice $r = \tilde\beta r_*$ coincide with, or at least approximate, the ones of the original nonlocal system. The correspondence between both systems is not injective because we have required to impose a further constraint (\ref{eqConstraint}). Since the localized system (\ref{LocalizedAction5D}) is second-order in spacetime components, we only need two initial conditions for each field $\Phi$ and $\chi$ instead of an infinite number of them. In particular, we need $\Phi(r,t_0,\bm{x})$, $\dot{\Phi}(r,t_0,\bm{x})$, $\chi(r,t_0,\bm{x})$, and $\dot{\chi}(r,t_0,\bm{x})$ at all $r$ and later evaluated at the special point with $r=\tilde\beta r_*$. However, since we have imposed (\ref{eqConstraint}) by hand, we have that the fields $\chi$ and $\Phi$ are not independent and the initial conditions $\chi(r,t_0,\bm{x})$ and $\dot{\chi}(r,t_0,\bm{x})$  at all values of $r$ are not independent from the ones of $\Phi$, so that we end up needing only two initial conditions for the field $\Phi(r,x)$ to specify the solution of nonlocal dynamics.

Once we have established the equivalence between the nonlocal system and one slice of the localized system, we may proceed to count the degrees of freedom of the theory. Using the Hamiltonian formalism \cite{Calcagni:2018lyd}, one may show that the field $\chi(r,x)$ is a ghost mode and the associated Hamiltonian is unbounded from below in the `five-dimensional' system. However, in the physical slice $r = \tilde\beta r_*$, this field disappears from the spectrum since its propagation is constrained by (\ref{eqConstraint}). Therefore, the absence of this ghost mode in the nonlocal four-dimensional system leaves us with only one propagating degree of freedom $\phi(x)$ that only requires two initial conditions $\phi(t_0,\bm{x})$ and $\dot{\phi}(t_0,\bm{x})$, plus the knowledge of the corresponding solution in the local system at $r_*=0$ (see below).

\subsubsection{Solutions}

One might wonder why we have not set $\tilde\beta = 0$ in the first place. This choice plays a special role in finding actual solutions. In fact, to construct these solutions, one needs to specify a \emph{seed} of the $\Phi$ field and then, using the diffusion equation, let this solution diffuse to the physical slice $r = \tilde\beta r_*$. The most natural choice for this seed is the solution to the local system, i.e., setting $r_* = 0$ in (\ref{nonlocalsys}), such that
\begin{equation}
    \Phi (0,t,\bm{x}) = \phi_{\rm local}(t,\bm{x}) \: .
\end{equation}
Thus, the solution $\phi_{\rm sol}(t,\bm{x})$ to the nonlocal system can be built out of $\phi_{\rm local}(t,\bm{x})$ through a diffusion via the expression
\begin{equation}
    \phi_{\rm sol}(x) = \Phi (\tilde\beta r_*,x) =  e^{\tilde\beta r_* \Box}\Phi(0,x) = \int_{-\infty}^{+\infty}\frac{d^4k}{(2\pi)^4}e^{-ik\cdot x}e^{-\tilde\beta r_* k^2}\Tilde{\phi}_{\rm local}(k) \: ,\label{EDMsolutions}
\end{equation}
where $\Tilde{\phi}_{\rm local}(k)$ is the Fourier transform of the solution $\phi_{\rm local}$ of the local system, i.e.,
\begin{equation}
    \Box \phi - V'(\phi) = 0\label{localeom} \: ,
\end{equation}
so that plugging the formal solution (\ref{EDMsolutions}) into the equation of motion (\ref{nonlocaleom}) one finds the value for $\beta$.

To conclude, we cite two main reasons for the choice of $\phi_{\rm local}$ as the seed of the diffusion.
\begin{enumerate}
    \item One expects to recover the solution of the local system when taking the limit $r_* \to 0$, and that is exactly what (\ref{EDMsolutions}) guarantees.
    \item Typically, diffusion does not alter the asymptotic behaviour of the field,
    \begin{equation}
        \lim_{x \to \pm \infty}\phi_{\rm local}(x) = \lim_{x \to \pm \infty}\phi_{\rm sol}(x) \: ,
    \end{equation}
    so that we may take the often known $\phi_{\rm local}$ as a guide to construct the nonlocal solutions $\phi_{\rm sol}$.
\end{enumerate}

\subsubsection{Solving a paradox}

At this point, one may be puzzled about the fate of the infinitely many initial conditions (\ref{InfiniteIC}) expected in the nonlocal model. Technically, it is clear that they are given by
\begin{equation}\label{inicon}
\phi^{(n)}(t_0,\bm{x})=\Phi^{(n)}(\tilde\beta r_*,t_0,\bm{x}) \quad \quad \forall n \in \mathbb{N} \: ,
\end{equation}
where $(n)$ is the $n$-th time derivative and the right-hand side is known after $\tilde\beta$ is determined by an algebraic equation. We started with infinitely many initial conditions that we encoded as one initial condition for the localized diffusing field $\Phi$. However, from the point of view of the nonlocal four-dimensional system one still needs infinitely many conditions and, for different initial conditions, there should be different available solutions. How is it possible that this infinite number of initial values have been reduced to two?

The problem with these questions is that they rely on the false premise that one \emph{could} solve the four-dimensional Cauchy problem if one knew infinitely many initial conditions. But this is not possible because it leads to the paradox we mentioned above: If one knows all the initial conditions, one already knows the solution. The diffusion method explains the paradox. From the solution $\phi_{\rm local}(x)=\Phi(0,x)$ of the local ($r_*=0$) four-dimensional system, via equation (\ref{EDMsolutions}) one \emph{does} know the solution $\Phi(\tilde\beta r_*,x)$ (except the value of $\tilde\beta$, easily found), so that one \emph{can} compute all the initial conditions from (\ref{inicon}). This knowledge, which looks to be needed `in advance' and therefore unattainable from the nonlocal four-dimensional perspective, is a simple consequence of the diffusion equation from the localized five-dimensional perspective. In other words, the Cauchy problem of the nonlocal system is \emph{defined} by the Cauchy problem of the localized diffusing system: this is the essence of the diffusion method. Without this definition, one is stuck with the impossibility of knowing a priori infinitely many initial conditions and, to the best of our knowledge, no alternative way out has been devised in the presence of interactions.\footnote{In contrast, it has been known since the early days that, for linear nonlocal equations of motion, one can apply nonlocal field redefinitions and reduce the dynamics to a local one \cite{Pais:1950za,Barnaby:2007ve}.}  

As a cautionary note, the diffusion method vastly constrains all possible solutions to the nonlocal system but, due to the lack of injectivity of the map between the four-dimensional nonlocal system and the localized one with the extra direction, it also does not exclude the existence of other physical solutions to the nonlocal system not obtainable using this procedure. In other words, we do not know whether the diffusion method covers the whole space of admissible solutions. To date, there are no counter-examples indicating such a possibility.

\section{Nonlocal classical gravity} \label{secNLG}

Having shown how healthy nonlocal operators may affect the UV behaviour of the propagator as well as the way the diffusion method allows one to make sense of the initial conditions and the physical modes of the theory, we generalize this approach to gravity and derive the equations of motion of this theory, as well as some immediate consequences for Ricci-flat spacetimes. We also comment on the way nonlocal gravity addresses the singularity problem when dealing with black holes.

\subsection{Action and equations of motion}

NLQG aims at solving the obstacles that Stelle's gravity (\ref{StelleAction}) experiences in order to be considered a complete theory of quantum gravity. As we have discussed previously, the presence of higher-derivative terms turns out to be a problem at the classical and quantum level because of the violation of unitarity. These higher-derivative terms are quadratic in the curvature operators $\mathcal{R}=R_{\mu \nu \rho \sigma}, R_{\mu \nu},R$ and their dependence on derivatives of the metric is
\begin{equation*}
    \mathcal{R} \sim \partial^2 g_{\mu \nu}, \partial g_{\mu \nu}^2 \quad \implies \quad \mathcal{R}^2 \sim (\partial^2 g_{\mu \nu})^2, (\partial g_{\mu \nu})^4, \partial^2 g_{\mu \nu}\partial g_{\mu \nu}^2 \: .
\end{equation*}
As we will show in section \ref{SecNLQG} at the tree level, the use of nonlocal operators in the action preserves unitarity, leading to a ghost-free theory. However, before we jump into quantum grounds, we formulate the classical theory of nonlocal gravity, whose action is given by
\begin{equation}
    S_{\rm NLG} =  \frac{1}{2 \kappa^2} \int d^4x \sqrt{-g}[ R +  R \gamma_0 (\Box) R +  R_{\mu \nu} \gamma_2(\Box) R^{\mu \nu}   +  R_{\mu \nu \rho \sigma} \gamma_4(\Box) R^{\mu \nu \rho \sigma} ] \: , \label{NLGActionGeneral}
\end{equation}
where $\gamma_0(\Box)$, $\gamma_2(\Box)$, and $\gamma_4(\Box)$ are the form factors of this theory. 

For the purposes of this section, it is enough to choose a particular set of form factors,
\begin{equation}
    \gamma_2(\Box) = - 2 \gamma_0(\Box) \equiv \gamma(\Box) \quad \quad \gamma_4(\Box) = 0 \: .
\end{equation}
With this choice, (\ref{NLGActionGeneral}) becomes
\begin{equation}
    S_{\rm NLG} = \frac{1}{2 \kappa^2}\int d^4 x \sqrt{-g} [ R + G_{\mu \nu} \gamma (\Box) R^{\mu \nu}] \label{simplifiedNLGActionF} \: .
\end{equation}

To derive the equations of motion of the action (\ref{simplifiedNLGActionF}), one may either vary directly the action or consider an auxiliary field that does not introduce additional degrees of freedom on-shell and coincides with the original action. Let us recall both approaches.

\subsubsection{Direct computation}

By varying the action (\ref{simplifiedNLGActionF}) in the presence of matter, i.e., $S = S_{\rm NLG} + S_{\rm m}$,
one finds the equations of motion \cite{Calcagni:2018lyd} (see also \cite{Koshelev:2013lfm,Biswas:2013cha})
\begin{equation}
\begin{split}
    \kappa^2 T_{\mu \nu} \:= \: &e^{-r_* \Box} G_{\mu \nu} - \frac{1}{2}g_{\mu \nu}G_{\rho \sigma} \gamma(\Box)R^{\rho \sigma} + 2 G^{\rho}_{(\mu}\gamma(\Box)G_{v)\rho} +g_{\mu \nu}\nabla^{\rho}\nabla^{\sigma}\gamma(\Box) G_{\rho \sigma} \\ & - 2\nabla^{\rho}\nabla_{(\mu} \gamma(\Box) G_{\nu) \rho} + \frac{1}{2}(G_{\mu \nu} \gamma (\Box) R + R \gamma(\Box) G_{\mu \nu}) + \Theta_{\mu \nu}(R_{\rho\sigma}, G^{\rho \sigma}) \label{EOMsNLG} \: ,
\end{split}
\end{equation}
where we introduced the energy-momentum tensor associated to the matter fields
\begin{equation}
    T_{\mu \nu} = - \frac{2}{\sqrt{-g}}\frac{\delta S_{\rm m}}{\delta g^{\mu\nu}} \: ,
\end{equation}
and for Wataghin form factor 
\begin{equation}
\Theta_{\mu \nu}(R_{\rho\sigma}, G^{\rho \sigma}) = - \int_0^{r_*}dq \Tilde{\Theta}_{\mu \nu}[e^{-q \Box}R_{\rho\sigma}, \frac{e^{-(r_* - q)\Box}-1}{\Box}G^{\rho \sigma}] \: .
\end{equation}
The tensor $\Tilde{\Theta}_{\mu \nu}$ can be split into two parts, one symmetric and one anti-symmetric with respect to the arguments $A$ and $B$, $\Tilde{\Theta}_{\mu \nu} = \Tilde{\Theta}_{\mu \nu}^{\text{sym}} + \Tilde{\Theta}_{\mu \nu}^{\text{antisym}}$, given by
\begin{equation}
    \Tilde{\Theta}_{\mu \nu}^{\text{sym}}(A_{\rho \sigma},B^{\rho\sigma}) = - \nabla_{\mu}A_{\rho \sigma}\nabla_{\nu}B^{\rho\sigma} + \frac{1}{4}g_{\mu \nu} \nabla_{\tau}(A_{\rho \sigma}\nabla^{\tau}B^{\rho \sigma} + B^{\rho \sigma} \nabla_{\tau}A_{\rho \sigma}) \: ,
\end{equation}
\begin{equation}
\begin{split}
    \Tilde{\Theta}_{\mu \nu}^{\text{antisym}}(A_{\rho \sigma},B^{\rho\sigma}) \:= \:&\frac{1}{4}g_{\mu \nu} \nabla_{\tau}(A_{\rho \sigma} \nabla^{\tau}B^{\rho \sigma} - B^{\rho \sigma} \nabla^{\tau}A_{\rho \sigma}) + \nabla_{\rho}(A_{\mu \sigma} \nabla^{\rho} B^{\sigma}_{\nu} - B^{\sigma}_{\nu}\nabla^{\rho}A_{\mu \sigma}) \\ & + \nabla_{\sigma}(B_{\mu \rho}\nabla_{\nu}A^{\rho \sigma} - A^{\rho \sigma}\nabla_{\nu}B_{\mu \rho}) + \nabla_{\rho}(A_{\mu\sigma}\nabla_{\nu}B^{\sigma\rho} - B^{\sigma\rho}\nabla_{\nu}A_{\mu\sigma}) \:.
\end{split}
\end{equation}
Here indices $\mu$ and $\nu$ are implicitly symmetrized. Notice that we recover Einstein's equations if we set $\gamma (\Box) = 0$.

\subsubsection{Auxiliary field}

Alternatively, we can derive the equations of motion introducing an auxiliary rank-2 symmetric tensor $\phi_{\mu \nu}$ in the action \cite{Calcagni:2018lyd}:
\begin{equation}
    \Tilde{S}[g,\phi] = \frac{1}{2\kappa^2}\int d^4x \sqrt{-g}\left[ R + (2R_{\mu \nu} - \phi_{\mu\nu} + \frac{1}{2}g_{\mu\nu}\phi)\gamma(\Box)\phi^{\mu \nu}\right] \label{AuxAction} \:,
\end{equation}
where $\phi = \phi^{\alpha}_{\alpha}$, and the respective equations of motion for both fields:
\begin{equation}
    \frac{\delta \Tilde{S}}{\delta g_{\mu\nu}} = 0 \:, \quad \quad \quad  \frac{\delta \Tilde{S}}{\delta \phi_{\mu\nu}} = 0 \:,
\end{equation}
are \cite{Calcagni:2018lyd}
\begin{equation}
\begin{split}
    \kappa^2 T_{\mu \nu} \:= \:& G_{\mu \nu} + \Box \gamma(\Box)\phi_{\mu \nu} - \frac{1}{2}g_{\mu \nu}X_{\rho \sigma} \gamma(\Box)\phi^{\rho \sigma} + 2 \phi^{\rho}_{(\mu}\gamma(\Box)\phi_{v)\rho} +g_{\mu \nu}\nabla^{\rho}\nabla^{\sigma}\gamma(\Box) \phi_{\rho \sigma} \\ & - 2\nabla^{\rho}\nabla_{(\mu} \gamma(\Box) \phi_{\nu) \rho} - \frac{1}{2}(\phi_{\mu \nu} \gamma (\Box) \phi + \phi \gamma(\Box) \phi_{\mu \nu}) + \Theta_{\mu \nu}(X_{\rho\sigma}, \phi^{\rho \sigma}) \: ,
\end{split}
\end{equation}
\begin{equation}
    \phi_{\mu\nu} = G_{\mu \nu} \label{PhiOS} \: ,
\end{equation}
where we have defined
\begin{equation}
    X_{\mu \nu} = 2 R_{\mu \nu} - \phi_{\mu \nu}+ \frac{1}{2}g_{\mu \nu} \phi \label{Xdef} \: .
\end{equation}
Note that we can calculate the trace of (\ref{PhiOS}) on-shell,
\begin{equation}
    \phi = G = - R\,, \quad \quad X_{\mu\nu} = R_{\mu \nu} \: ,
\end{equation}
and we recover (\ref{EOMsNLG}).

\subsection{Diffusion method for nonlocal gravity} \label{SecDEMNLG}

Similarly to what we did in section \ref{sectionDEM}, one can show that this nonlocal formulation of gravity does not bring ghost modes to the particle spectrum of the theory and the initial conditions problem is solved consistently. We take the minimal operator (\ref{WataghinFF}), referring the reader to \cite{Calcagni:2018gke} for the case of asymptotically polynomial form factors. 

The action that one must take into account is given by \cite{Calcagni:2018lyd}
\begin{equation}
    S[\Phi,g,\chi,\lambda] = \frac{1}{2 \kappa^2}\int d^4 x dr \sqrt{-g} ( \mathcal{L}_R +  \mathcal{L}_{\Phi} +  \mathcal{L}_{\chi} +  \mathcal{L}_{\lambda} ) \: ,
\end{equation}
with
\begin{equation}
     \mathcal{L}_R = R(r) \: ,
\end{equation}
\begin{equation}
     \mathcal{L}_{\Phi} = - \int_0^{r_*}ds \left[ 2\Tilde{\mathcal{R}}_{\mu \nu}(r) - \Phi(r) + \frac{1}{2}g_{\mu\nu}(r) \Phi(r)  \right]\Phi^{\mu \nu}(r-s) \: ,
\end{equation}
\begin{equation}
     \mathcal{L}_{\chi} = - \int_0^{r_*}ds\int_0^s \chi_{\mu\nu}(r-q) (\partial_{r'}-\Box)\Phi^{\mu\nu}(r') \: ,
\end{equation}
\begin{equation}
     \mathcal{L}_{\lambda} = \lambda_{\mu\nu}\partial_r g^{\mu\nu}(r) \label{LagMult} \: ,
\end{equation}
where we have omitted the $x$-dependence in all fields, and $\Tilde{\mathcal{R}}_{\mu \nu}$ is the Ricci tensor associated to $g_{\mu\nu}(r)$, that on-shell will become $r$-independent. In fact
\begin{equation}
    \frac{\delta S}{\delta \lambda_{\mu\nu}} = 0 \quad \implies \quad \partial_r g^{\mu\nu} = 0 \quad \implies \quad g_{\mu\nu}(r,x) = g_{\mu\nu}(x) \: .
\end{equation}
From this localized Lagrangian, we can obtain the equations of motion:
\begin{equation}
    (\partial_r - \Box)\Phi_{\mu\nu} = 0 \: ,
\end{equation}
\begin{equation}
    (\partial_r - \Box)\chi_{\mu\nu} = 0 \: ,
\end{equation}
\begin{equation}
    \int_0^{r_*}ds[ X_{\mu\nu}(r-s) + X_{\mu\nu}(r+s) - 2 \Tilde{\mathcal{R}}_{\mu\nu}(r-s) + \chi_{\mu\nu}(r-s) - \chi_{\mu\nu}(r+s)] = 0 \label{eqCCC} \: ,
\end{equation}
\begin{equation}
\begin{split}
    \kappa^2 T_{\mu\nu} \:= \: & G_{\mu\nu} - \int_0^{r_*}ds \{-\frac{1}{2}g_{\mu\nu}X_{\alpha\beta}(r)\Phi^{\alpha\beta}(r-s) + 2 \Phi_{\sigma (\mu}(r)\Phi_{\nu)}^{\sigma}(r-s) \\ & + \Box \Phi_{\mu\nu}(r-s) + g_{\mu\nu}\nabla^{\sigma}\nabla^{\tau}\Phi_{\sigma\tau}(r-s) - e \nabla^{\sigma}\nabla_{(\mu}\Phi{\nu)\sigma}(r-s) \\ &  - \frac{1}{2}[\Phi_{\mu\nu}(r)\Phi(r-s) + \Phi(r)\Phi_{\mu\nu}(r-s)] \\ &  - \int_0^s dq \bar{\Theta}_{\mu\nu}[\chi_{\sigma \tau}(r-q),\Phi^{\sigma\tau}(r+q-s)] \} \: , \label{LongEOM}
\end{split}
\end{equation}
where $X_{\mu\nu}$ is
\begin{equation}
    X_{\mu\nu}(r) = 2\Tilde{\mathcal{R}}_{\mu\nu}(r) - \Phi_{\mu\nu}(r) + \frac{1}{2}g_{\mu\nu}(r)\Phi(r) \: .
\end{equation}
From the first two equations, we see that the auxiliary fields $\Phi_{\mu\nu}$ and $\chi_{\mu\nu}$ propagate via a diffusion equation. Similarly to what we did in the scalar field theory, we may impose by hand a constraint at the slice $r = \tilde\beta r_*$ analogous to (\ref{eqConstraint}),
\begin{equation}
    \chi_{\mu\nu}(\tilde\beta r_*) = X_{\mu\nu}(\tilde\beta r_*) = R_{\mu\nu} \quad \implies \quad \Phi_{\mu\nu}(\tilde\beta r_*) = \phi_{\mu\nu} = G_{\mu\nu} \label{EqConstraintGrav} \: ,
\end{equation}
which satisfies (\ref{eqCCC}). Using the diffusion equation for $\Phi_{\mu\nu}$, we have
\begin{equation}
    -\int_0^{r_*}ds \Phi_{\mu\nu}(r-s) = \frac{e^{-r_*\Box}-1}{\Box}\Phi_{\mu\nu}(r) = \gamma (\Box) \Phi_{\mu\nu}(r) \: ,
\end{equation}
so that at the physical slice $\tilde\beta r_*$ we recover the equations of motion (\ref{EOMsNLG}) of the nonlocal theory.

The equivalence between the localized five-dimensional system  at the slice $r = \tilde\beta r_*$ and the nonlocal four-dimensional theory with the minimal form factor can be generalized to other exponential-monomial form factors such as Krasnikov's. However, when dealing with asymptotically polynomial form factors one has to follow a different approach \cite{Calcagni:2018gke} where a diffusion-like equation is implemented in a more sophisticated way.

\subsubsection{Initial conditions, degrees of freedom and absence of ghosts}

In the gravitational case, we have to introduce three additional fields: $\Phi_{\mu\nu}$, $\chi_{\mu\nu}$, and $\lambda_{\mu\nu}$. The latter was simply used to implement the condition $\partial_r g_{\mu\nu}$ and does not show any dynamics by itself. On the other hand, the field $\chi_{\mu\nu}$ has a similar origin that the scalar field $\chi$ introduced in (\ref{ScalarLagrangianChi}). In a similar vein, we have constrained by hand this $\chi_{\mu\nu}$ by the expression (\ref{EqConstraintGrav}), so that $\chi_{\mu\nu} = R_{\mu\nu} \sim \partial g_{\mu\nu}^2, \partial^2 g_{\mu\nu}$ as well as $\Phi_{\mu\nu}$ on-shell by (\ref{PhiOS}), so that they do not become additional propagating modes, since their diffusion is frozen in the $\tilde\beta r_*$ slice.

From this relation between $\chi_{\mu\nu}$ and the Ricci tensor, we conclude that for the minimal/Wataghin form factor we only need four initial conditions 
\begin{equation}
    g_{\mu\nu}(t_0,\bm{x}),\quad \dot{g}_{\mu\nu}(t_0,\bm{x}),\quad \ddot{g}_{\mu\nu}(t_0,\bm{x}),\quad \dddot{g}_{\mu\nu}(t_0,\bm{x}),
\end{equation}
instead of an infinite number of them. This number of initial values can increase for other types of form factors but it remains finite.

We also have to count the physical degrees of freedom of this theory, and to do so we have to find the propagating independent components of the tensorial fields of our equations. First of all, let us recall how many  physical degrees of freedom contains the graviton: since it is a rank-2 symmetric tensor, in four dimensions it has 10 independent components, but gauge invariance coming from diffeomorphism invariance reduces them to 6. Finally, the contracted Bianchi identities $\nabla_{\mu}G^{\mu\nu} = 0$ reduce by 4 this amount, giving rise to only 2 independent components of the graviton.

On the other hand, for the auxiliary field $\phi_{\mu\nu} = G_{\mu\nu}$ on-shell, only the Bianchi identities apply, so that it contains 6 degrees of freedom. In conclusion, we have in total $6+2 = 8$ propagating degrees of freedom, result that coincides with the counting in Stelle's gravity \cite{Stelle:1976gc}. The main difference here is that nonlocal gravity does not contain any ghost field so that the theory is stable. In fact, one may follow the perturbative procedure described in \cite{Stelle:1976gc} to show that, expanding the Lagrangian at second order in the perturbation/graviton field $h_{\mu\nu}$ defined as
\begin{equation}
    g_{\mu\nu} = \eta_{\mu\nu} + 2\kappa h_{\mu\nu} \label{eqGraviton} \: ,
\end{equation}
the ghost disappears from the particle spectrum of the nonlocal theory. This Lagrangian at second order has the following expression for a generic form factor (\ref{APFF}) \cite{Calcagni:2018gke}:
\begin{equation}
    \mathcal{L}^{(2)} = \mathcal{L}_E(h_{\mu\nu}) + 3 \phi \frac{\Box}{1-e^{- \text{H}(\Box)}}\phi - \frac{1}{2}\psi_{\mu\nu}(\eta^{\mu\rho}\eta^{\nu\sigma}-\eta^{\mu\nu}\eta^{\rho\sigma})\frac{\Box}{1-e^{- \text{H}(\Box)}}\psi_{\rho\sigma} \label{eq89} \:,
\end{equation}
where $\psi_{\mu\nu}$ is the traceless part of $\phi_{\mu\nu}$, and $\phi$ is its trace, and $\mathcal{L}_E$ is the Einstein's linearized Lagrangian given by \cite{Gravitation}
\begin{equation}
    \mathcal{L}_E(h_{\mu\nu}) = \frac{1}{2}h_{\mu\nu}\Box h^{\mu\nu} - \frac{1}{2}h\Box h + h^{\mu\nu}\partial_{\mu}\partial_{\nu}h - h^{\mu\nu}\partial_{\rho}\partial_{\nu}h^{\rho}_{\mu} \: .
\end{equation}
From these expressions, one sees that the second and the third term in (\ref{eq89}) give rise to the mass of the additional modes $\phi$ and $\psi_{\mu\nu}$. Stelle's gravity can be interpreted as a truncated expansion of the minimal/Wataghin form factor, in particular, one recovers Stelle's particle spectrum in the limit
\begin{equation}
   \frac{e^{ \text{H}(\Box)}}{\gamma (\Box)} = \frac{\Box}{1-e^{- \text{H}(\Box)}} \simeq \frac{\Box}{ \text{H}(\Box)} \simeq 1 \label{eq91}  \: ,
\end{equation}
giving rise to propagators with the wrong sign, i.e., ghost fields.

Nevertheless, the nonlocal operator (\ref{eq91}) cannot be truncated in our gravitational theory. In particular, the propagator of the fields $\psi_{\mu\nu}$ and $\phi$ will be proportional to $\gamma e^{-\text{H}}$ and, by definition of the form factors, it will never display new extra poles in the theory.

\subsubsection{Solutions}

In analogy with what we did in the scalar field case, we can construct exact or approximate solutions of the nonlocal system using as a seed of the diffusion equation the local system, i.e., taking $\gamma(\Box) = 0 \implies r_* = 0$, such that the equations of motion are given by the Einstein's equations
\begin{equation}
    \kappa^2T_{\mu\nu} = G_{\mu\nu} \quad \implies \quad \chi_{\mu\nu}(0,x) = R_{\mu\nu}^{\rm local}(x), \quad \Phi_{\mu\nu}(0,x) = G_{\mu\nu}^{\rm local}(x) \: ,
\end{equation}
where $R_{\mu\nu}^{\rm local}$ and $G_{\mu\nu}^{\rm local}$ are built with a solution $g_{\mu\nu}^{\rm local}$ of the local system. From these local solutions, one can diffuse to the physical slice $r \to \tilde\beta r_*$. However, one must note that in a curved spacetime, the integral representation of the kernel (\ref{MinkFTFF}) is no longer correct since the functions $\text{exp}(\pm i k \cdot x)$ are not eigenfunctions of the Laplace--Beltrami operator. In this case, one has to find two eigenfunctions of $\Box$ and write (\ref{MinkFTFF}) as their linear superposition \cite{Calcagni:2007ru}, as explained in section \ref{SectionAppendix}.

\subsection{Analysis and properties of NLG}\label{SecRicciFlat}

Once derived the equations of motion of the minimal nonlocal gravitational theory (\ref{simplifiedNLGActionF}) we may explore some of its classical properties. In general, one expects that finding analytic solutions to these equations will not be possible. However, the structure of the equations (\ref{EOMsNLG}) allows us to say a few things about its solutions when considering Ricci-flat spacetimes. We can also prove the stability of these solutions under small perturbations as we have recently done in (\ref{eqGraviton}) on Minkowski but including more general backgrounds where the additional degrees of freedom $\psi_{\mu\nu}$ and $\phi$ can propagate. Lastly, we check that the stability of these backgrounds can be generalized to any perturbative order.

\subsubsection{Ricci-flat spacetimes} \label{sectionRF}

The first question we may ask ourselves is whether it is possible to export the solutions of Einstein's gravity to NLG. The very structure of the equations of motion (\ref{EOMsNLG}) tells us that it is possible, in particular, if we consider Ricci-flat spacetimes or vacuum solutions of Einstein's gravity,
\begin{equation}
    R_{\mu\nu} = 0 \quad \implies \quad G_{\mu\nu} = 0 \quad\implies \quad T_{\mu\nu}=0 \: ,
\end{equation}
then the equations of motion of NLG are automatically satisfied so that they constitute valid solutions of the nonlocal theory, such as Minkowski, Schwarzschild or Kerr spacetime.

NLG aspires to avoid the singularities arising in GR, that tell us that the theory breaks down in some region of spacetime. This problem was initially identified by Schwarzschild \cite{Schwarzschild:1916uq} and Hilbert \cite{Hilbert:1915tx} when they studied the metric nowadays taking the former's name:
\begin{equation}
    ds^2 = - \left(1 - \frac{2GM}{r}\right)dt^2 + \left(1 - \frac{2GM}{r}\right)^{-1}dr^2 + r^2 (d \theta^2 + \sin^2 \theta d \phi^2) \: . \label{SchwMetric}
\end{equation}
This metric blows up at $r= 2GM$ and $r = 0$, but one easily shows that, while the first singularity is caused by an inappropriate choice of coordinates, the second is an $O(1/r)$ curvature singularity that cannot be removed by a suitable coordinate transformation. Furthermore, in the non-relativistic limit, the Schwarzschild metric reproduces Newtonian gravity after linearizing the metric
\begin{equation}
    ds^2 = - [ 1 + 2 \Phi (\bm{x})]dt^2 + [1 - 2\Phi (\bm{x})]dr^2+ r^2(d\theta^2 + \sin^2\theta d\phi^2) \: ,
\end{equation}
such that the Newtonian potential $\Phi (\bm{x})$ satisfies the Poisson's equation
\begin{equation}
    \partial_i \partial^i \Phi = \delta^{(3)}(\bm{x}) \: .
\end{equation}

\subsubsection{Stability}\label{stabi}

We say that a background solution $g^{(0)}_{\mu\nu}$ is stable against linear perturbations $h_{\mu\nu}$ if the metric $g_{\mu\nu} = g^{(0)}_{\mu\nu} + h_{\mu\nu} $ does not blow up when it solves the vacuum equations of motion, i.e., if the perturbation $h_{\mu\nu}$ remains small throughout the dynamical evolution. From this definition, one may prove that Minkowski \cite{Lindblad:2004ue} and Schwarzschild (\ref{SchwMetric}) \cite{Regge:1957td} spacetimes are stable in GR.

Furthermore, when $\gamma_4=0$ in the action (\ref{NLGActionGeneral}), we also have that Schwarzschild \cite{Calcagni:2017sov} and Minkowski \cite{Briscese:2018bny} spacetimes are stable against linear perturbations in NLQG, and that this result can be generalized to all Ricci-flat spacetimes \cite{Calcagni:2018pro} and all perturbative orders \cite{Briscese:2019rii}. Note, however, that this argument is based on the results exposed in section \ref{sectionRF} in which we see the direct link between Einstein's gravity and NLG in Ricci-flat spacetimes, so that stability in GR is inherited by the nonlocal theory.

To prove the latter statement, it is convenient to introduce the Lichnerowicz operator $\Delta_L$, that acting on a rank-2 symmetric field $\psi_{\mu\nu}$ is defined as
\begin{equation}
    \Delta_L  \psi_{\mu\nu}=2 R^{\rho}_{\mu\nu\tau}\psi^{\tau}_{\rho} + R_{\mu\rho}\psi^{\rho}_{\nu} + R_{\sigma \nu}\psi^{\sigma}_{\mu} - \Box \psi_{\mu\nu} \: , 
\end{equation}
and acting on a scalar it is simply given by $\Delta_L \psi = - \Box \psi$.

Along with this new operator we can define a theory formally equivalent to (\ref{NLGActionGeneral}) but with the substitution $-\Box \to \Delta_L$ everywhere, namely
\begin{equation}
    \mathcal{L} = R + G_{\mu\nu}\gamma(\Delta_L)R^{\mu\nu} \label{LichLagrangian} \: .
\end{equation}
Clearly, this theory and the minimal theory (\ref{simplifiedNLGActionF}) share the same renormalization properties on Minkowski spacetime and, for simplicity, we use the Lagrangian (\ref{LichLagrangian}) to prove stability order by order. The proof of the stability for the minimal theory is more involved \cite{Briscese:2018bny} but, since the perturbative expansion around Minkowski spacetime is the same for both theories, the result reviewed here is applicable to the theory of our interest.

The equations of motion of this nonlocal theory are given by \cite{Calcagni:2018pro}
\begin{equation}
    e^{H(\Delta_L)}G_{\mu\nu} + \mathcal{O}_{\mu\nu}(\mathcal{R}^2) = 0 \: , \label{eqLichnerowiczEOM}
\end{equation}
where $\mathcal{O}_{\mu\nu}(\mathcal{R}^2)$ denotes terms at least quadratic in the Ricci tensor $R_{\mu\nu}$. This form of the equations of motion is compatible with the solution $R_{\mu\nu} = 0$. One can directly substitute $R_{\mu\nu} = G_{\mu\nu}$ and acting $e^{-H(\Delta_L)}$ on both sides of (\ref{eqLichnerowiczEOM}), one finds the nested expression
\begin{equation}
    G_{\mu\nu} = e^{-H(\Delta_L)}\mathcal{O}_{\mu\nu}(\mathcal{G}^2) \label{eqStabilityLich} \: ,
\end{equation}
where $\mathcal{O}_{\mu\nu}(\mathcal{G}^2)$ denotes terms at least quadratic in the Einstein tensor $G_{\mu\nu}$, 
\begin{equation}
    \mathcal{O}_{\mu\nu}(\mathcal{G}^2) = [\mathcal{G} \mathcal{O}\mathcal{G}]_{\mu \nu} +  \mathcal{O}_{\mu\nu}(\mathcal{G}^3) \label{Gexpansion} \: ,
\end{equation}
being $\mathcal{O}$ an operator that acts on $G_{\mu\nu}$ in both sides. Taking the perturbative expansion
\begin{equation}
    g_{\mu\nu} = \sum_{n=0}^{\infty}\epsilon^n h_{\mu\nu}^{(n)} , \quad \quad h_{\mu\nu}^{(0)} = g_{\mu\nu}^{(0)} \: ,
\end{equation}
with $\epsilon \ll 1$, the Einstein tensor $G_{\mu\nu}$ to all orders is
\begin{equation}
    G_{\mu\nu}(g_{\mu\nu}) = \sum_{n=0}^{\infty}\epsilon^n G_{\mu\nu}^{(n)} \: ,
\end{equation}
where $G_{\mu\nu}^{(0)} = 0$ because the background metric is Ricci-flat. Also, the exponential operator is abstractly written as
\begin{equation}
     e^{-H(\Delta_L)} = \sum_{n=0}^{\infty}\epsilon^n S^{(n)} \: .
\end{equation}
Since $\mathcal{G}(g_{\mu\nu}) \sim \epsilon$, we may neglect cubic terms in the expasion (\ref{Gexpansion}) and, eliminating the tensorial structure of the previous expressions, we can write (\ref{eqStabilityLich}) order by order as
\begin{equation}
     \mathcal{G}^{(n)} = \sum_{h=0}^n \sum_{k=0}^h \sum_{q=0}^{k}S^{(n-h)}\mathcal{G}^{(h-k)}\mathcal{O}^{(k-q)}\mathcal{G}^{(q)} \: .
\end{equation}
By recursion, one obtains that $\mathcal{G}^{(n)} = 0 $ $\forall n \in \mathbb{N}$. For instance, taking $n=1$ one gets
\begin{equation}
   \mathcal{G}^{(1)} = S^{(0)}(\mathcal{G}^{(1)}\mathcal{O}^{(0)}\mathcal{G}^{(0)} + \mathcal{G}^{(0)}\mathcal{O}^{(1)}\mathcal{G}^{(0)} + \mathcal{G}^{(0)}\mathcal{O}^{(0)}\mathcal{G}^{(1)}) +  S^{(1)} \mathcal{G}^{(0)}\mathcal{O}^{(0)}\mathcal{G}^{(0)} = 0\,.
\end{equation}
Therefore, one concludes that the solutions of the nonlocal theory (\ref{LichLagrangian}) are stable at any order if they are stable in Einstein's theory. 

This result also holds for the theory (\ref{simplifiedNLGActionF}) \cite{Briscese:2018bny} and has immediate consequences on the discussion about the additional 6 degrees of freedom $\psi_{\mu\nu}$ and $\phi$ of section \ref{SecDEMNLG}. There, we showed that they did not propagate in Minkowski spacetime, but now we realize that this is true on any Ricci-flat metric. In particular, on these backgrounds the stability of the nonlocal theory is inherited from Einstein's theory, in which there are no additional propagating degrees of freedom other than the two corresponding to the graviton. Consequently, the ghost modes $\psi_{\mu\nu}$ and $\phi$ are not dynamical fields and the theory is unitary.

\subsection{Smoothing out singularities}

To wrap up this part on the classical theory, we recall a key property of weakly nonlocal operators that consists in \emph{smearing} the singularities, which makes this theory so appealing in order to deal with the singularity problem arising in GR. After showing how nonlocality is able to smooth out classical divergences, we briefly discuss implications for black holes.

\begin{figure}[H]
    \centering
    \includegraphics[width = 0.7\textwidth]{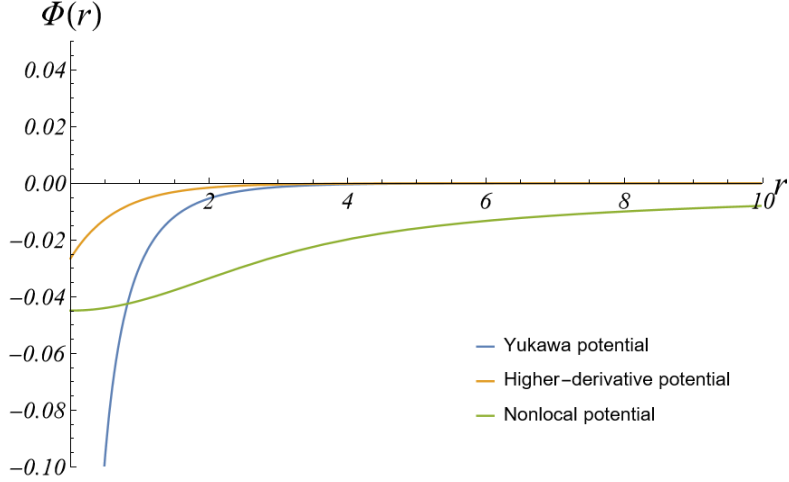}
    \caption{$\Phi(r)$ for the three cases described in the text taking $l_* = 1$, $m=m_1=1$, and $m_2 = 2$. In blue the standard Yukawa potential, in orange the higher-derivative (four-derivative) case and in green the nonlocal (Wataghin form factor) case.}
    \label{fig:Potential}
\end{figure}

Let us start by some preliminaries that illustrate the manner form factors avoid singularities. First of all, consider the Poisson equation for the static gravitational potential $\Phi(\bm{x})$ in three-dimensional flat space,
\begin{equation}\label{poieq}
    \gamma(\Box) \Phi(\bm{x}) = \delta^{(3)}(\bm{x}) \: ,
\end{equation}
where $\Box = \partial_i \partial^i$. There are three cases of interest: the classical second-derivative operator, higher-derivative (HD) operators and nonlocal form factors (figure \ref{fig:Potential}).
\begin{itemize}
    \item {\textbf{Standard Yukawa potential.}} In this case, $\gamma(\Box) = \Box - m^2$ and one can calculate $\Phi(\bm{x})$ using spherical coordinates:
    \begin{equation*}
       (\Box - m^2) \Phi(\bm{x}) = \delta^{(3)}(\bm{x}) \quad \implies \quad \Phi(\bm{x}) = - \int \frac{d^3\bm{k}}{(2\pi)^3}\frac{e^{+i\bm{k}\cdot \bm{x}}}{k^2+m^2} \: ,
    \end{equation*}
    whose value is given by
    \begin{equation}
        \Phi (r) = - \frac{e^{-mr}}{4\pi r} \: ,
    \end{equation}
    which is divergent in the limit $r=|\bm{x}| \to 0$. When the mass parameter $m=0$, then we have the standard Newtonian potential falling off as $1/r$.
    \item {\textbf{Quartic in derivatives local HD form factor.}} In this case, $\gamma(\Box) = - (\Box - m_1^2)( \Box - m_2^2)$ and
        \begin{equation*}
       -(\Box - m_1^2)(\Box - m_2^2) \Phi(x) = \delta^{(3)}(\bm{x}) \: \implies \: \Phi(\bm{x}) =  -\int \frac{d^3\bm{k}}{(2\pi)^3}\frac{e^{i\bm{k}\cdot \bm{x}}}{(k^2+m^2_1)(k^2+m^2_2)}\: .
    \end{equation*}
    Splitting the integral as
    \begin{equation*}
        \frac{1}{(k^2+m^2_1)(k^2+m^2_2)} = \frac{1}{m_1^2-m_2^2}\left[ \frac{1}{k^2+m_2^2} - \frac{1}{k^2+m_1^2} \right] \:,
    \end{equation*}
 we can make use of the results of the Yukawa potential to write
    \begin{equation}
        \Phi(r) = \frac{1}{m_1^2-m_2^2} \frac{e^{-m_1r}-e^{-m_2r}}{4 \pi r} \: . \label{HDsmearing}
    \end{equation}
    When taking the limit $r \to 0$, $\Phi(r)$ remains finite,
    \begin{equation}
        \Phi(0) = - \frac{1}{4\pi (m_1+m_2)} \: .
    \end{equation}
    \item {\textbf{Wataghin form factor.}} In this case, we choose $\gamma(\Box) = e^{-l_*^2\Box}\Box$, so that the potential is given by
    \begin{equation*}
         e^{-l_*^2\Box}\Box\Phi(\bm{x}) = \delta^{(3)}(\bm{x}) \quad \implies \quad \Phi(\bm{x}) =  \int \frac{d^3\bm{k}}{(2\pi)^3}\frac{e^{+i\bm{k}\cdot\bm{x} - l_*^2k^2}}{k^2}\: .
    \end{equation*}
    The final result is
    \begin{equation}
        \Phi (r) = - \frac{1}{4\pi r} \text{erf}\left( \frac{r}{2l_*}\right) \: , \label{NLsmearing}
    \end{equation}
    where $\text{erf}$ is the error function. When taking the limit $r \to 0$, one finds
    \begin{equation}
        \Phi(r\to 0 ) = -\frac{1}{4\pi^{3/2}l_*} \: ,
    \end{equation}
    which is finite.
\end{itemize}
In conclusion, using nonlocal form factors allows one to remove the classical divergences of gravity by smearing the gravitational source, in the same way Wataghin \cite{Wataghin:1934ann} gave a finite radius to point-like particles using nonlocal operators.

Some comments about these results are in order here. First, we have not derived the Yukawa (Newtonian) potential from the full equations of motion using the diffusion method but we assumed the modified Poisson equation (\ref{poieq}). Second, the analysis was done for the non-relativistic linear Yukawa potential, without non-linearities and using the Fourier transform. However, this procedure might not give all solutions, since a hidden condition is that $\Phi$ fall off at spatial infinity. Third, in section \ref{sectionRF} we showed that Ricci-flat spacetimes are also solutions of the equations of motion of NLG, which means that there do exist singular black holes such as those described by Schwarzschild and Kerr metrics.

There are two ways in which NLG can approach the singularity problem based on which type of nonlocal theory (\ref{NLGActionGeneral}) we adopt. 
\begin{itemize}
    \item If we omit the Riemann-Riemann term and set $\gamma_4(\Box) = 0$, as we did in sections \ref{sectionRF} and \ref{stabi}, then the results of Ricci-flat spacetimes apply and the singularities of the Schwarzschild and Kerr black holes are transferred to the nonlocal theory. Therefore, this nonlocal gravitational theory is not capable to tame the singularities of GR at the classical level. Fortunately, conformal invariance solves the problem at the quantum level \cite{Modesto:2016max}. The status of Ricci-flat solutions as physical is also not completely established. In fact, one can argue that, since astronomical black holes are formed by gravitational collapse, there must be matter inside the event horizon, so that vacuum solutions to Einstein's equations are not valid at the singularity $r=0$. Thus, as is well known in GR, Schwarzschild's metric (\ref{SchwMetric}) does not describe all the spacetime but only a patch of it, and the matter distribution associated with Schwarzschild should be computed solving Einstein's equations in the sense of distribution. This has been done for GR \cite{Balasin:1993fn,Steinbauer:2006qi} but not yet in NLG, so that we do not really know that the singularity of the metric consistently matches a delta-like matter distribution at the center of the black hole.
    \item Setting $\gamma_4(\Box) \neq 0$, the Ricci-flat results are no longer valid since in the equations of motion we have terms involving the Riemann tensor, which, in general, is non-vanishing even for Ricci-flat spacetimes. In this case, nonlocality can make the singularities disappear already at the classical level \cite{Biswas:2005qr,Biswas:2011ar,Biswas:2012bp}. Indeed, the higher-derivative and nonlocal operators studied in this review smooth out point-like singularities \cite{Giacchini:2016xns,Modesto:2014eta,Buoninfante:2018xiw}, as seen in equations (\ref{HDsmearing}) and (\ref{NLsmearing}). Also, it has been proven that the presence of a Riemann-Riemann term in the Lagrangian forbids Schwarzschild-type singularities with $\Phi(r)\sim 1/r^{\alpha}$ for $\alpha > 1$ \cite{Koshelev:2018hpt,Buoninfante:2018xiw}. However, a problem of the nonlocal theory with the Riemann-Riemann tensor terms is that, in general, it is difficult to find exact solutions.
\end{itemize}

\section{Nonlocal quantum scalar field theory}

Having studied the classical theory of nonlocal interactions (or nonlocal kinetic terms, which is the same as we saw in section \ref{noloca}), we now focus on how we quantize this theory and how unitarity is preserved at the quantum level. First of all, we review some basics about power.counting renormalization as well as the derivation of the unitarity bound. Subsequently, we analyze the way nonlocal form factors alter the prescription of local theories to deal with momentum integrals of the quantum theory and how we can employ Efimov's analytic continuation to build a meaningful quantum theory. To conclude, we introduce the Cutkosky rules used to verify the perturbative unitarity of the nonlocal scalar theory.

\subsection{Power-counting renormalizability} \label{sectionPowerCounting}

Renormalization is a property of a quantum theory that tells us whether our theory may calculate physical quantities, namely if it is predictive. In practice, the renormalization procedure is related to the way we manage to deal with the infinities that can appear in momentum integrals in a quantum theory. There are many ways in which we can regularize a theory, i.e., express these infinities, such as the cut-off scheme, dimensional regularization, and so on. Once the theory is regularized, one can check whether it can be renormalized. In this section, we approach the renormalizability of a theory via the so-called \emph{power-counting renormalizability}, which is a useful criterion to know how badly the scattering amplitudes of the theory will be divergent.

First, let us review how this criterion works for the local scalar field theory. In general, the Lagrangian of these theories can be written as
\begin{equation}
    \mathcal{L}_{\phi} = \sum_n \lambda_n \mathcal{O}^{(n)}(\phi) \: ,
\end{equation}
where $\mathcal{O}^{(n)}$ denotes an interacting term for the scalar field $\phi$ with energy dimension $n$ and $\lambda_n$ are the bare coupling constants. After renormalization, these couplings acquire a dependence on the energy scale of the given process and are not constant in general.

Depending on the way scalar fields interact in $\mathcal{O}^{(n)}$ the energy dimensionality of the coupling $\lambda_n$ will change. In particular, if $D$ is the topological dimension of the spacetime, we may distinguish three different cases.
\begin{itemize}
    \item $ n < D$. In this case, $[\lambda_n] > 0$, and the coupling will decrease as we increase the energy, so that this kind of operator is called \emph{relevant}, since it becomes important in the IR regime.
    \item $ n = D$. In this case, $[\lambda_n] = 0$, and the coupling will not depend on the energy scale, so that this kind of operator is called \emph{marginal}.
    \item $ n > D$. In this case, $[\lambda_n] < 0$, and the coupling will grow as we go to higher and higher energies, so that this kind of operator is called \emph{irrelevant}, since it becomes important in the UV regime. These constitute the non-renormalizable couplings of a theory.\footnote{In $D=4$ dimensions, the allowed renormalizable interactions at the Lagrangian level are $\phi^2$,$\phi^3$, and $\phi^4$, since interactions of the kind $\phi^n$ for $n \geqslant 5$ are non-renormalizable. The Standard model of particle physics is built out of field operators of maximal energy dimension 4.}
\end{itemize}
Once identified the type of operators that our scalar theory might have, one applies the renormalization procedure, that allows us to reabsorb the divergences of scattering amplitudes. In particular, the scattering amplitudes $\mathcal{M}$ involving some of the previous operators diverge. However, one can reabsorb these infinities order by order by including higher-order operators at the Lagrangian level, so that the divergent part of the amplitude is canceled out. In the case of irrelevant operators, one needs to add an infinite number of higher-order operators, which results in a non-renormalizable theory. In this context, one says that the theory is \emph{power-counting renormalizable} if the interactions satisfy $[\lambda_n] \geqslant 0$.

In the nonlocal case, one must be more careful with these arguments, since there are infinitely many interactions that involve derivatives. We introduce the \emph{superficial degree of divergence} $\omega(\mathcal{F})$ of a Feynman diagram $\mathcal{F}$ as a criterion to classify a given theory as renormalizable or non-renormalizable. This number characterizes a particular Feynman diagram and it encodes the energy dependence of the scattering amplitude associated with it when computing the corresponding integrals up to a given cut-off scale $\Lambda$ that will be taken to infinity:
\begin{equation}
    \mathcal{M} \propto \Lambda^{\omega(\mathcal{F})} \:.
\end{equation}
From this expression, one concludes that:
\begin{itemize}
    \item $\omega(\mathcal{F}) < 0 \quad \implies \quad \mathcal{F}$ is superficially convergent.
    \item $\omega(\mathcal{F}) = 0 \quad \implies \quad \mathcal{F}$ diverges logarithmically.
    \item $\omega(\mathcal{F}) > 0 \quad \implies \quad \mathcal{F}$ is superficially divergent.
\end{itemize}
Actually, the power-counting renormalization is not a necessary condition for the renormalizability of the theory itself. For instance, there are finite scattering amplitudes whose superficial degree of divergence is positive, such as the one-loop 4-photon diagram in QED that cancels out by Furry's theorem despite $\omega(\mathcal{F}) > 0$ \cite{Peskin:1995ev}.

Therefore, we may use this quantity to qualitatively classify a Feynman diagram as divergent or not, but always keeping in mind that there could be some symmetry mechanism in the system that ameliorates the divergent behaviour of a given scattering amplitude. In contrast, power-counting renormalizability is a sufficient condition for renormalizability, so that, if a theory is found to be power-counting renormalizable, then explicit calculations of scattering amplitudes will only confirm this result.

\subsection{Unitarity}

In a QFT, one of the most important quantities is the so-called scattering amplitude, often denoted by $\mathcal{M}$. The square of this object, $|\mathcal{M}|^2$, is used to calculate cross-sections of a particular decay and characterizes the probability with which this process can take place. Therefore, the calculation of $|\mathcal{M}|^2$ can be directly compared with experimental data to validate or not a theory.

From a classical point of view, we briefly mentioned in section \ref{secOstro} that the existence of higher-order derivatives in a theory gives rise to unstable modes that, in turn, lead to a spontaneous decay of the vacuum. In this sense, we say that unitarity is broken and, since our theory cannot reproduce a physical universe, it must be ruled out. In the quantum regime, this condition of unitarity can be easily encoded in the S-matrix, defined as the matrix $\hat{S}$ that connects the initial state \emph{a} and the final state \emph{b} in a particular decay. Mathematically, it can be written as
\begin{equation}
     \mathcal{S}_{a \rightarrow b} =\bra{b}\hat{S}\ket{a} .
\end{equation}
By conservation of probability, one has that the S-matrix is a unitary matrix that satisfies 
\begin{equation}
    \hat{S}^{\dagger}\hat{S} = \id \label{eqUnitarity} \:.
\end{equation}
It is usually convenient to split the S-matrix into 2 parts: one that describes the free theory in which no interaction is taken into account and another whose main purpose is to describe interactions. One writes this splitting as
\begin{equation}
    \hat{S} = \id + i \hat{T} \label{eqSplit}\:,
\end{equation}
where the matrix $\hat{T}$ has been introduced.

\subsubsection{Optical theorem} \label{sectionOpticalTheorem}

Inserting the splitting (\ref{eqSplit}) into the unitarity condition (\ref{eqUnitarity}), one obtains

\begin{equation}
    -i(\hat{T}- \hat{T}^{\dagger}) = \hat{T}^{\dagger}\hat{T} \: .
\end{equation}
We can act on this expression with an initial state $\ket{i}$ and a final state $\ket{f}$ and define the amplitudes $\mathcal{T}_{fi} = \bra{f} \hat{T} \ket{i}$. Using this notation, we have that
\begin{equation}
     -i(\bra{f}\hat{T}\ket{i}- \bra{f}\hat{T}^{\dagger}\ket{i}) = -i(\mathcal{T}_{fi} - \mathcal{T}_{if}^*) = \bra{f}\hat{T}^{\dagger}\hat{T}\ket{i} \label{eqOpt1}\:.
\end{equation}
Considering a theory invariant under the reflection $x^{\mu} \rightarrow - x^{\mu}$, one has that $\hat{T}$ is symmetric, and inserting a completeness relation in between $\hat{T}$ and $\hat{T}^{\dagger}$ in the right-hand side of (\ref{eqOpt1}), one obtains
\begin{equation}
    -i(\mathcal{T}_{fi} - \mathcal{T}_{fi}^*) = 2 \text{Im}\mathcal{T}_{fi}
    = \sum\mathop{}_{\mkern-5mu n} \mathcal{T}_{nf}^* \mathcal{T}_{ni} \: .
\end{equation}
This result is known as the \emph{optical theorem} and, as we will see shortly, it will be of crucial importance in order to test the unitarity of a nonlocal theory. Besides, in the particular case of the forward scattering, i.e., $i = f$, one finds that in a theory with no ghosts \cite{Peskin:1995ev}
\begin{equation}
    2 \text{Im}T_{ii} = \sum\mathop{}_{\mkern-5mu n} |\mathcal{T}_{ni}|^2 > 0 \quad \implies \quad \text{Im}T_{ii} > 0 \label{eqBound1} \: .
\end{equation}
Furthermore, if one factorizes out the momentum dependence that always appears in these matrix elements and instead works in terms of the scattering amplitude $\mathcal{M}$ defined as
\begin{equation}
\mathcal{T}_{ij} = (2 \pi)^4 \delta^{(4)}(\text{P}_T) \mathcal{M}_{ij}  \label{eqUnitSplit}\: ,
\end{equation}
 one has that (\ref{eqBound1}) becomes
 \begin{equation}
     |\mathcal{M}_{ii}|  \geqslant \text{Im}\mathcal{M}_{ii} = \frac{1}{2} \sum\mathop{}_{\mkern-5mu n} |\mathcal{M}_{ni}|^2 \geqslant \frac{1}{2} |\mathcal{M}_{ii}|^2 \quad \implies \quad |\mathcal{M}_{ii}| \leq 2 \: .
 \end{equation}
In addition, using this result for $|\mathcal{M}_{ii}|$ and considering a general state $j$,  one has that
 \begin{equation}
      2 \geqslant |\mathcal{M}_{ii}|  \geqslant \text{Im}\mathcal{M}_{ii} = \frac{1}{2} \sum\mathop{}_{\mkern-5mu n} |\mathcal{M}_{ni}|^2 \geqslant  \frac{1}{2} |\mathcal{M}_{ij}|^2 \quad \implies \quad |\mathcal{M}_{ij}| \leq 2 \: .
 \end{equation}
 This \emph{unitarity bound} must be satisfied by any quantum field theory that preserves unitarity. In section \ref{sectionTreeLUnitarity}, we will show that, indeed, this unitarity bound is satisfied in NLQG.
 
\subsection{Lorentzian and Euclidean momentum space}

As briefly mentioned in section \ref{subsectionSummary}, we can define a quantum field theory either in Lorentzian momentum space or in Euclidean momentum space. An analytic continuation $k_E = -i k^0$ of the energy component of the momenta, making it purely imaginary, connects the two formulations by a Wick rotation. Since the propagator usually has poles in $k^0$, which are identified with particle modes of the theory, one extends the domain of $k^0$ to the complex plane using an analytic continuation and use the Feynman prescription $i\epsilon$ to displace these poles from the real axis. As a result of this displacement in the complex plane, one may use Cauchy's theorem to calculate the integrals that usually appear in any quantum theory, namely,
\begin{equation}
    I = \int_{-\infty}^{+\infty} dk^0 f(k^0) \label{CommonIntegralQFT} \: .
\end{equation}
 In this way, one relates both momenta prescription to characterize the same theory
 and one can calculate quantities in Euclidean momentum space and then apply an inverse (clockwise) Wick rotation to return to the theory in Lorentzian momentum. This is the standard procedure when dealing with local scalar field theories.

However, in nonlocal field theories this equivalence is not available since the contributions of the arcs at infinity, in general, do not vanish because of the additional momentum dependence in the propagator. In fact, most nonlocal propagators diverge in some quadrant of the complex plane when going to infinity.

As showed in Figure \ref{fig:Euclidean}, it is desirable to define the nonlocal theory in Euclidean momenta since the UV behaviour of the nonlocal form factors of our interest is divergent, the propagator is highly suppressed in this regime and, as a consequence, the potential divergences of the quantum theory are reduced. Since we do not have Wick rotation at our disposal to convert to Lorentzian signature, we need to implement an alternative procedure called \emph{Efimov analytic continuation}. Before we introduce this prescription, we talk about an important feature of nonlocal theories that allows us to redefine our fields without inducing observable consequences.

\subsubsection{Nonlocal field redefinition}

One aspect of nonlocal field theories that we have not approached yet is a possible nonlocal redefinition of the fields, so that in some way we may get an equivalent local theory. To be more precise, given the nonlocal theory
\begin{equation}
    \mathcal{L} = \frac{1}{2}\phi \gamma (\Box) \phi - V(\phi) \label{LagrangianEquivTheorem} \: ,
\end{equation}
is it possible to redefine our field as $\phi\to\Tilde{\phi}(\phi)$ such that the theory becomes local without modifying the physics? This is in general possible under some particular assumptions. However, to show it explicitly we make use of the following form factor:
\begin{equation}
    \gamma(\Box) = e^{\text{H}(\Box)}(\Box - m^2) \: .
\end{equation}
There exist \emph{equivalence theorems} \cite{Kamefuchi:1961sb,Chisholm:1961tha} proving that, under nonlinear local field redefinition at the Lagrangian level, one obtains the same scattering amplitudes than the original theory; namely, under a transformation of the form
\begin{equation}
    \Tilde{\phi}(\phi) = \phi + f(\phi, \nabla \phi) \: ,
\end{equation}
the observable quantities remain invariant. However, although it was shown that this equivalence also holds for certain nonlocal redefinitions \cite{Bergere:1975tr}, in general a nonlocal field redefinition might not lead to an equivalent theory. Here we focus only on non-singular and non-zero nonlocal form factors, so that a field redefinition
\begin{equation}
    \Tilde{\phi}(\phi) = e^{\frac{1}{2}\text{H}(\Box)}\phi \label{NLFRD} \: ,
\end{equation}
does leave the particle spectrum invariant, since no new poles appear in the propagator. Therefore such a nonlocal transformation leads to an equivalent theory \cite{Bergere:1975tr}. \\ \\
Under the redefinition (\ref{NLFRD}), (\ref{LagrangianEquivTheorem}) transforms into
\begin{equation}
    \mathcal{L} = \frac{1}{2}\Tilde{\phi}(\Box - m^2) \Tilde{\phi} - V[e^{-\frac{1}{2}\text{H}(\Box)}\Tilde{\phi}], \label{NLSFTtransferred}
\end{equation}
so that in the free field theory ($V = 0$) the nonlocal redefinition (\ref{NLFRD}) leads to a local scalar field theory. In general, we can always transfer all the nonlocality to the potential, which usually is a polynomial of third or higher order, leading to nonlocal interactions and vertices. As a result of this, we may use the standard propagator of the local theory given by
\begin{equation}
G(k^2) = \frac{1}{- k^2-m^2 + i \epsilon} \label{CommonPropagator} \: .
\end{equation}

\subsubsection{Unitarity} \label{sectionCUTKOSKY}

As we said, we cannot apply a Wick rotation to establish the equivalence of the quantum theory expressed in Euclidean and Lorentzian momentum space. However, Efimov worked out a consistent way to identify both formulations \cite{Efimov:1967dpd} which is based on an analytic continuation to the complex plane of the time-like component of both the internal and external momenta in a Feynman diagram. In this prescription, one carries out the explicit calculations integrating the time-like component of the momentum along the imaginary axis and afterwards, the external momentum is analytically continued back to its real value. The main difference with respect to the traditional Wick rotation is that the way to return to the real values of the external momenta is not achieved by something looking like a rigid rotation but by a specific and more complicated deformation of the integration contour in the complex plane.

The optical theorem stated in section \ref{sectionOpticalTheorem} can be generalized via the \emph{Cutkosky rules} \cite{Cutkosky1960} which can be applied to the nonlocal Lagrangian (\ref{NLSFTtransferred}) in order to prove the unitarity of the theory. These Cutkosky rules are based on the possibility to cut a Feynman diagram into two pieces so that we may decompose the full diagram into the sum of intermediate states. As we have showed in section \ref{sectionOpticalTheorem}, to prove unitarity we need to focus on the imaginary part of the scattering amplitude $\mathcal{M}$. In this context, the Cutkosky rules state that, if the theory is unitary, then we can replace the internal propagators by a delta function \cite{Cutkosky1960}:
\begin{equation}
    \frac{i}{k^2+m^2-i\epsilon} \quad \to \quad 2\pi i \delta (k^0)\delta(k^2+m^2) \: .
\end{equation}
In conclusion, to prove perturbative unitarity of the nonlocal theory, one can calculate its scattering amplitudes $\mathcal{M}$ explicitly and show that the imaginary part is given by the Cutkosky rules. Since these rules hold when the theory is unitary, one shows that the nonlocal theory does not violate unitarity \cite{Briscese:2018oyx}.


\section{Nonlocal quantum gravity} \label{SecNLQG}

The last part of this chapter is dedicated to show how nonlocal gravity solves the obstacles that Einstein's gravity poses when one tries to quantize it. First of all, we expose the problems that GR has in the context of renormalization as well as the way Stelle's theory overcomes them. Then we tackle the problem of ghosts. We have already seen in section \ref{SecDEMNLG} the classical instabilities of this modified gravity caused by the appearance of kinetic terms in the linearized action with the wrong sign. Here we show how Stelle's gravity violates the unitarity bound derived in section \ref{sectionOpticalTheorem}, as well as how nonlocality provides an answer to the unitarity problem.

Secondly, we can see that no breaking of causality is induced in this theory by looking at Saphiro's time delay. We also study the superficial degree of divergence $\omega(\mathcal{F})$ of the theory and see that $\omega(\mathcal{F})$ is positive in the one-loop case so that we focus on this particular setup. After showing how to achieve renormalizability for the nonlocal theory, we highlight the asymptotic freedom that the theory exhibits.

In order to quantize the gravitational field, we split the metric field into a background field, that will be chosen to be flat-spacetime, and a perturbation that is identified with the graviton. Under the decomposition (\ref{eqGraviton}), one rewrites Einstein's action (\ref{EHAction}) in terms of the perturbation $h_{\mu\nu}$ to obtain a canonical kinetic term proportional to $(\partial h)^2$, as well as higher-order interactions of the graviton. The resulting Lagrangian contains graviton interactions with coupling constants whose energy dimension is negative, leading to a non-renormalizable operator studied through power-counting arguments.

Early analyses in the 1960s by Feynman \cite{Feynman:1963ax} and DeWitt \cite{DeWitt:1967} showed that the one-loop renormalizability of the theory required the addition of Faddeev--Popov ghosts that, unlike the ordinary ghosts that plague higher-derivative gravity, do not propagate in external legs. Furthermore, in the 1970s, 't Hooft and Veltman \cite{tHooft:1974toh} extensively studied all the one-loop divergences of Einstein's action and proved that pure gravity without matter, was renormalizable at the one-loop level, while it was not when considering its coupling to matter. The final explicit calculation to confirm the non-renormalization of the theory was performed by Goroff and Sagnotti \cite{Goroff:1985sz,Goroff:1985th}, who showed that two-loop divergences of pure gravity were unavoidable.

From then on, many gravitational theories and approaches have been proposed but, for our purposes, we now comment on some important features of Stelle's gravity (\ref{StelleAction}), whose coupling constants are dimensionless and it results in a satisfactory renormalization of the theory \cite{Stelle:1976gc,Stelle:1977ry}. 

\subsection{Graviton propagator}

In order to construct the quantum theory, we begin by computing the graviton propagator of the nonlocal gravity in $D=4$ dimensions: 
\begin{equation}
    \mathcal{L} = \frac{1}{2 \kappa^2}\left[ R + R \gamma_0(\Box)R + R_{\mu\nu} \gamma_2(\Box)R^{\mu\nu} \right] \label{eqGenNLQG} \: ,
\end{equation}
where we have applied the Gauss--Bonnet theorem to get rid of the Riemann-Riemann tensor term and the form factors $\gamma_0(\Box)$ and $\gamma_2(\Box)$ are defined as (\ref{APFF}) in the following way:
\begin{equation}
    \gamma_0(\Box) =  \frac{e^{\text{H}_0(\Box)}-1}{2\Box}\: , \quad \quad \gamma_2(\Box) = -\frac{e^{\text{H}_2(\Box)}-1}{\Box} \label{PropFF} \: .
\end{equation}
Notice the dimensionless couplings $\left[ \kappa^{-2} \gamma_{0,2}\right] = 0$.

Expanding this nonlocal action to second order in the background-plus-graviton decomposition (\ref{eqGraviton}), one obtains \cite{Accioly:2002tz}
\begin{equation}
\begin{split}
    \mathcal{L}^{(2)} \:= \:&  \:\mathcal{L}^{(2)}_{E} + \frac{1}{2} \{  \Box h_{\mu\nu} \gamma_2(\Box) h^{\mu\nu} - \partial_{\mu}A^{\mu}\gamma_2(\Box)\partial_{\nu}A^{\nu} - F^{\mu\nu}\gamma_2(\Box)F_{\mu\nu} + \\ & + (\partial_{\mu}A^{\mu}-\Box h)\left[ \gamma_2(\Box) + 4 \gamma_0(\Box)\right](\partial_{\nu}A^{\nu}-\Box h) \} \: ,
\end{split} \label{eqLinearizedL}
\end{equation}
where we have neglected total derivatives and defined the quantities $A^{\mu} = \partial_{\nu}h^{\mu\nu}$ and $F_{\mu\nu} = 2 \: \partial_{[\mu}A_{\nu]}$, and $\Box = \eta_{\mu\nu}\partial_{\mu}\partial_{\nu}$ at this order in the expansion.

Diffeomorphism invariance constitutes the gauge symmetry of the gravitational theory and it is manifested by the invariance of the Lagrangian (\ref{eqLinearizedL}) under the coordinate transformation \cite{poisson_will_2014}
\begin{equation}
    x^{\mu} \to x^{\mu} + \zeta^{\mu} \quad \rightarrow \quad h_{\mu\nu} \to  h_{\mu\nu} - 2 \: \partial_{(\mu}\zeta_{\nu)} \: .
\end{equation}
Together with this local symmetry, one introduces a gauge-fixing term at the Lagrangian level given by the harmonic or De Donder gauge \cite{Stelle:1976gc}
\begin{equation}
    \mathcal{L}_{\text{gf}} = - \frac{1}{\xi} A_{\mu} \: \omega (\Box) \: A^{\mu} \: ,
\end{equation}
where $\omega(\Box)$ is a dimensionless weight function \cite{Stelle:1976gc,Buchbinder:1992rb} and $\xi$ is the gauge-fixing parameter. Then, the complete Lagrangian is given by
\begin{equation}
    \mathcal{L}^{(2)} +  \mathcal{L}_{\text{gf}} = \frac{1}{2}h^{\mu\nu}\mathcal{O}_{\mu\nu\rho\sigma}h^{\rho\sigma} \: ,
\end{equation}
where the inverse of $\mathcal{O}_{\mu\nu\rho\sigma}$ is the graviton propagator
\begin{equation}
    \bra{0}\hat{T}\lbrace h_{\mu\nu}(x)h_{\rho\sigma}(y)\rbrace \ket{0} = i \mathcal{O}^{-1}_{\mu\nu\rho\sigma} \: .
\end{equation}
Following the procedure of \cite{Accioly:2002tz} we find the explicit expression of the graviton propagator in terms of the Barnes--Rivers operators \cite{doi:10.1063/1.1704335,Rivers1964LagrangianTF}
\begin{equation*}
\begin{split}
     P_{\mu\nu\rho\sigma}^{(0)} = \: & \frac{1}{3}\theta_{\mu\nu}\theta_{\rho\sigma} \: , \\
     P_{\mu\nu\rho\sigma}^{(1)} = \: & \theta_{\mu(\rho}\omega_{\sigma)\nu} + \theta_{\nu(\rho}\omega_{\sigma)\mu} \: , \\
     P_{\mu\nu\rho\sigma}^{(2)} = \: & \theta_{\mu(\rho}\theta_{\sigma)\nu} - \frac{1}{3}\theta_{\mu\nu}\theta_{\rho\sigma} \: , \\
    \bar{P}_{\mu\nu\rho\sigma}^{(0)} =  \: &\omega_{\mu\nu}\omega_{\rho\sigma} \: , \\
\end{split}
\end{equation*}
where $\theta_{\mu\nu} = \eta_{\mu\nu}-\frac{k_{\mu}k_{\nu}}{k^2}$ represents the transversal vector projection operator and $\omega_{\mu\nu} =  \frac{k_{\mu}k_{\nu}}{k^2}$ is the longitudinal vector projection operator. Using these operators and their mathematical properties, one derives the graviton propagator of the nonlocal theory (\ref{eqGenNLQG}) in momentum space \cite{Accioly:2002tz}
\begin{equation}
    \mathcal{O}^{-1} = -\frac{1}{k^2}\left[ \frac{P^{(2)}}{1-k^2\gamma_2(k^2)} - \frac{P^{(0)}}{2+4k^2 [3 \gamma_0(k^2) + \gamma_2(k^2)]} + \frac{\xi(2P^{(1)}+\bar{P}^{(0)})}{2 \omega(k^2)}\right] \: ,
    \label{eqGravitonPropagator}
\end{equation}
and using the convenient definition of the form factors (\ref{PropFF}) we find
\begin{equation}
    \mathcal{O}^{-1} =  \frac{1}{k^2}\left[ \frac{P^{(2)}}{e^{\text{H}_2(k^)}} - \frac{P^{(0)}}{2e^{\text{H}_0(k^2)}}\right] - \frac{\xi(2P^{(1)}+\bar{P}^{(0)})}{2k^2\omega(k^2)}  \label{GeneralPropagator} \: ,
\end{equation}
where we have omitted the index structure of the tensors. Assuming the minimal choice $\text{H}_0 = \text{H}_2 \equiv \text{H}$ (\ref{WataghinFF}), the expression (\ref{GeneralPropagator}) becomes \cite{Modesto:2014lga}
\begin{equation}
    \mathcal{O}^{-1} =  \frac{e^{-\text{H}}}{k^2}\left[ P^{(2)} - \frac{P^{(0)}}{2}\right] - \frac{\xi(2P^{(1)}+\bar{P}^{(0)})}{2k^2\omega(k^2)} \: ,
\end{equation}
where the first part of the propagator coincides with the one obtained in Einstein's gravity but with an inverse exponential that improves the UV convergence of the theory. Moreover, note the particular expression of the form factors (\ref{PropFF}) with the inverse Laplace--Beltrami operator that is canceled by the $k^2$ factors that go with the form factors in (\ref{eqGravitonPropagator}), as well as the generality of the propagator, that is valid even for a local theory with polynomials $\text{H}_0(\Box)=p_0(\Box)$ and $\text{H}_2(\Box)=p_2(\Box)$ satisfying $p_0(0) = p_2(0) = 1$.


\subsection{Tree-level unitarity} \label{sectionTreeLUnitarity}

Once introduced the graviton propagator of the nonlocal theory we are in position to prove the unitarity of the theory. In particular, we begin the analysis focusing on tree-level unitarity and finally we briefly recall how to prove the perturbative unitarity of the theory.

We have already mentioned the connection between unitarity and ghost modes at the classical level. However, when dealing with a quantum theory, the presence of ghosts in the particle spectrum is related to the order of poles of the propagator, namely, the theory is unitary if the corresponding propagator has only simple poles in $k^2+m^2 = 0$, such that the residues of the propagator are positive; a negative residue indicates violation of unitarity as one sees from (\ref{eq2poles}).

The unitarity test is performed when we couple the graviton to a general conserved stress-energy tensor $T^{\mu\nu}$ and check that the scattering amplitudes satisfy the unitarity bound (\ref{eqBound1}). In this approach, the linearized Lagrangian becomes
\begin{equation}
    \mathcal{L}_{hT} = \frac{1}{2}h^{\mu\nu}\mathcal{O}_{\mu\nu\rho\sigma}h^{\rho\sigma} - \sqrt{2}h_{\mu\nu}T^{\mu\nu} \: ,
\end{equation}
and one computes the scattering amplitude using the perturbative decomposition (\ref{eqSplit}) and (\ref{eqUnitSplit}) of a initial state $i$ to decay into a final state $f$ \cite{Accioly:2002tz}
\begin{equation}
    \bra{f}i\hat{T}\ket{i} = (2\pi)^4 \delta^{(4)}(P_T)i\mathcal{M}_{if} = (2\pi)^4 \delta^{(4)}(P_T)i^2 T^{\mu\nu}i\mathcal{O}^{-1}_{\mu\nu\rho\sigma}T^{\rho\sigma} \: .
\end{equation}
To find explicit expressions, one decomposes the tensor $T^{\mu\nu}$ in terms of the following independent vectors in momentum space:
\begin{equation}
    k^{\mu} = (k^0,k^i), \quad \Tilde{k}^{\mu}=(-k^0,k^i),\quad \epsilon_i^{\mu} = (0, \epsilon^i) \: ,
\end{equation}
for $i=1,2$, being $\epsilon^{\mu}_i$ orthogonal to $k^{\mu}$. The decomposition is done in a completely general way, and one writes
\begin{equation}
    T^{\mu\nu} = a k^{\mu}k^{\nu} + b \Tilde{k}^{\mu}\Tilde{k}^{\nu} + c^{ij}\epsilon^{(\mu}_i \epsilon^{\nu)}_j + d k^{(\mu}\Tilde{k}^{\nu)} + e^i k^{(\mu}\epsilon^{\nu)}_i+ f^i \Tilde{k}^{(\mu}\epsilon^{\nu)}_{i} \label{generalT} \: ,
\end{equation}
and imposing conservation of the stress-energy tensor in momentum space, i.e., $k_{\mu}T^{\mu\nu}=0$, one finds that $d=b=f^i=0$. If we want to compute $2 \text{Im}\mathcal{M}_{if}$, we need the imaginary part of
\begin{equation}
    \mathcal{M}_{if} = - T^{\mu\nu}i\mathcal{O}^{-1}_{\mu\nu\rho\sigma}T^{\rho\sigma} \: ,
\end{equation}
where the graviton propagator has been obtained in (\ref{GeneralPropagator}). The gauge-dependent part of the propagator vanishes when contracted with the stress-energy tensors, so that we can neglect this part and rewrite the propagator, including the $+i\epsilon$ prescription, as
\begin{equation}
    i\mathcal{O}^{-1} = \frac{i}{i\epsilon-k^2}\left[ \frac{P^{(2)}}{e^{\text{H}_2}} - \frac{P^{(0)}}{2e^{\text{H}_0}} \right],
\end{equation}
so that
\begin{equation}
    \mathcal{M}_{if} = - T^{\mu\nu}\frac{1}{i\epsilon-k^2}\left[ \frac{P^{(2)}}{e^{\text{H}_2}} - \frac{P^{(0)}}{2e^{\text{H}_0}} \right]_{\mu\nu\rho\sigma}T^{\rho\sigma} \: .
\end{equation}
To calculate the imaginary part of the previous expression, we start focusing on $\frac{1}{i\epsilon-k^2}$:
\begin{equation}
    \text{Im}\left( \frac{1}{i\epsilon-k^2} \right) = \frac{1}{2i}\left( \frac{1}{i\epsilon-k^2} - \frac{1}{-i\epsilon-k^2}\right) = \frac{\epsilon}{k^4 + \epsilon^2} \stackrel{\epsilon \to 0}{\longrightarrow} \pi \delta(k^2) \: .
\end{equation}
On the other hand, using the formul\ae\ of the Barnes--Rivers operators, one gets
\begin{equation}
    2 \text{Im}\mathcal{M}_{if} = 2 \left[ T_{\mu\nu}T^{\mu\nu} - \frac{T^2}{2}\right] \pi \delta(k^2)  \: ,
\end{equation}
where $T = T^{\mu}_{\mu}$ is the trace of the stress-energy tensor, and we have used the property $\text{H}_0 (0) = \text{H}_2 (0) = 0$. Finally, using the expression for $T^{\mu\nu}$ with $b=d=f^i=0$ along with the condition $k_{\mu}T^{\mu\nu} = 0$, one finds that
\begin{equation}
     2 \text{Im}\mathcal{M}_{if} = 2 \left[ (c^{ij})^2 - \frac{(c^{ii})^2}{2}\right]  \pi \delta(k^2) \: ,
\end{equation}
and since
\begin{equation}
c^{ij} =
\begin{pmatrix}
c^{11} & c^{12} \\ c^{21} & c^{22}
\end{pmatrix} \: ,
\end{equation}
we have that
\begin{equation}
    (c^{ij})^2 - \frac{(c^{ii})^2}{2} = \frac{1}{2}\left[(c^{11})^2 + (c^{22})^2\right] + (c^{12})^2 + (c^{21})^2.
\end{equation}
Thus, one concludes that the unitarity bound $2 \text{Im} \mathcal{M}_{if} > 0 $ is satified and the theory is unitary at the tree-level.

\subsubsection{Perturbative unitarity}

As we commented on section \ref{sectionCUTKOSKY}, perturbative unitarity of the nonlocal scalar theory is achieved by means of Efimov analytic continuation together with the verification of the Cutkosky rules. For NLQG, the procedure is analogous but with slightly different Cutkosky rules. The explicit proof of perturbative unitarity is tedious \cite{MarcoPiva,Briscese:2013lna} but, intuitively, on may reason that since the nonlocal theory only affects the Einstein's graviton propagator with the inclusion of the form factors, which do not produce extra poles, no unitarity violation is induced.

\subsection{Tree-level scattering amplitudes}

Once sketched the proof of the unitarity of the nonlocal theory, in this part we recall the calculation of some graviton scattering amplitudes at tree-level for both Stelle's gravity and NLQG using the convenient transverse traceless gauge for the graviton field. Furthermore, we show an important result that relates the scattering amplitudes of these theories at tree-level with the ones computed using Einstein's gravity \cite{Dona:2015tra}.

\begin{figure}[h]
\centering
 \includegraphics[width = 0.3\textwidth]{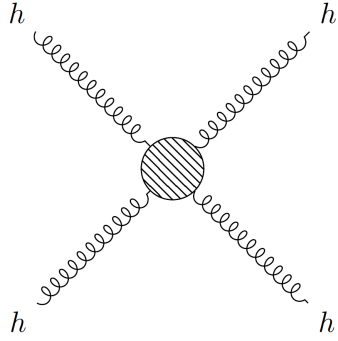}
    \caption{2-2 graviton scattering process.}
    \label{fig:22scattering}
\end{figure}

First, in order to calculate scattering amplitudes conveniently, we work in the transverse traceless gauge for the graviton field, so that any on-shell graviton in the external legs of the scattering process satisfies the gauge-fixing conditions
\begin{equation}
    \partial^{\mu}h_{\mu\nu}= 0 \quad \quad h = h^{\mu}_{\mu}=0 \: . \label{TransverseTracelessGauge}
\end{equation}
We now define the dimensionless graviton field $[ h] = 0$ as
\begin{equation}
    g_{\mu\nu} = \eta_{\mu\nu} + h_{\mu\nu} \label{Dimensionlessh} \: .
\end{equation}
In this gauge, we may decompose the polarization tensor $\epsilon_{\mu\nu}$ of the spin-2 graviton as a combination of the spin-1 photon polarization vectors $\epsilon_{\mu}$ as \cite{Gravitation}
\begin{equation}
    \epsilon_{\mu\nu}(p, \pm2) = \epsilon_{\mu}(p, \pm1)\epsilon_{\nu}(p, \pm1) \: ,
\end{equation}
with 
\begin{equation}
    \epsilon_{\mu}(p,\lambda)p^{\mu} = 0 \quad \quad \epsilon_{\mu}(\lambda)\epsilon^{\mu}(\lambda)=0 \: .
\end{equation}
We define the helicity amplitudes $F_{\lambda_3\lambda_4;\lambda_1\lambda_2}$ for a 4 graviton scattering 1 + 2 $\to$ 3 + 4 by the relation with the already introduced scattering amplitude $\mathcal{M}$ via
\begin{equation}
    F_{\lambda_3,\lambda_4;\lambda_1,\lambda_2} = \frac{i}{\mathcal{N}}\bra{\lambda_3,\lambda_4}\mathcal{M}\ket{\lambda_1,\lambda_2} \: ,
\end{equation}
where $\mathcal{N}$ is a normalization factor.
\\ \\
The introduction of these helicity amplitudes allows one to obtain four handy properties for this particular 2-2 scattering process, depicted in Figure (\ref{fig:22scattering}).

Imposing the individual discrete symmetries $C$, $P$, and $T$, together with Bose symmetry because of the bosonic character of the gravitons, one finds the following four properties of helicity amplitudes \cite{Grisaru:1975bx}:
\begin{itemize}
    \item $P$ symmetry: 
    \begin{equation}
        F_{\lambda_3,\lambda_4;\lambda_1,\lambda_2}  =  (-1)^{\lambda-\mu}F_{-\lambda_3,-\lambda_4;-\lambda_1,-\lambda_2} \: .
    \end{equation} 
    \item $T$ symmetry in the process $a + b \to a + b$:
    \begin{equation}
        F_{\lambda_3,\lambda_4;\lambda_1,\lambda_2}  =  (-1)^{\lambda-\mu}F_{\lambda_1,\lambda_2;\lambda_3,\lambda_4} \: .
    \end{equation}
    \item $C$ symmetry in the process $a + \bar{a} \to b + \bar{b}$:  \begin{equation}
         F_{\lambda_3,\lambda_4;\lambda_1,\lambda_2}  =  (-1)^{\lambda-4}F_{\lambda_3,\lambda_4;\lambda_2,\lambda_1} \: .
    \end{equation}
    \item Bose symmetry in the process $a + a \to b + c$:  \begin{equation}
        F_{\lambda_3,\lambda_4;\lambda_1,\lambda_2}  =  (-1)^{\lambda-\mu}F_{\lambda_4,\lambda_3;\lambda_2,\lambda_1} \: ,
    \end{equation} 
\end{itemize}
where $\lambda = \lambda_1 - \lambda_2$ and $\mu = \lambda_3 - \lambda_4$. Taking into account these properties, we find that out of 16 initial independent components, we are only left with 4:
\begin{equation*}
\begin{split}
        \mathcal{A}(++;++) \equiv F_{2,2;2,2}\,, \quad & \quad  \mathcal{A}(+-;+-) \equiv  F_{2,-2;2,-2}\,, \\
        \mathcal{A}(++;+-) \equiv F_{2,2;2,-2}\,, \quad & \quad  \mathcal{A}(++;--)\equiv  F_{2,2;-2,-2}\,. \label{4scatteringAmplitudes}
\end{split}
\end{equation*}

\subsubsection{Helicity amplitudes in local gravity}

For the action (\ref{StelleAction}), using the perturbative decomposition (\ref{Dimensionlessh}) one now computes the 3- and 4-point functions of the theory, expanding up to third and fourth order in the graviton $h_{\mu\nu}$. Besides, since we aim to compute the 2-2 scattering amplitudes, for the sake of simplicity we may assume two on-shell gravitons in the following derivations so that the gauge conditions (\ref{TransverseTracelessGauge}) may be imposed onto two out of three of the gravitons present in the 3-point functions and on the four gravitons present in the 4-point function. In particular, imposing on-shell conditions in the external gravitons determines that the linearized Ricci scalar and the determinant of the metric vanish,
\begin{equation}
    R^{(1)} = \eta^{\mu\nu}R_{\mu\nu}^{(1)} = \partial_{\alpha}\partial_{\beta}h^{\alpha\beta} - \Box h = 0 \: ,
\end{equation}
\begin{equation}
    \sqrt{-g}^{(1)} = \sqrt{-h}^{(1)} = 0 \: .
\end{equation}
Moreover, the linearized equations of motion imply that the Ricci tensor at linear order vanishes as well,
\begin{equation}
    \left(\frac{\delta S}{\delta g_{\mu\nu}}\right)^{(1)} = 0 \implies R_{\mu\nu}^{(1)} = \frac{1}{2}(-\partial_{\mu}\partial_{\nu}h + \partial_{\mu}\partial^{\alpha}h_{\alpha\nu} + \partial_{\nu}\partial^{\alpha}h_{\alpha\mu} - \Box h_{\mu\nu}) = 0 \: .
\end{equation}
Furthermore, using the Gauss--Bonnet theorem we may rewrite Stelle's action (\ref{StelleAction}) as
\begin{equation}
    S_{\rm Stelle} = \frac{1}{2\kappa^2}\int d^4x \sqrt{-g}(R + \gamma_0R^2 + \gamma_2R_{\mu\nu}R^{\mu\nu}) = S_{\text{E}} + \gamma_0 S_0 + \gamma_2 S_2 \: , \label{StelleSimplifiedAction}
\end{equation}
where $\gamma_0 = \alpha_1 - \alpha_3$ and $ \gamma_2 = \alpha_2 + 4 \alpha_3$.

\subsubsection{Equations of motion}

For the 3-point function, one needs to expand (\ref{StelleSimplifiedAction}) at second order and use the vanishing values for the Ricci Scalar, Ricci tensor and the trace of the graviton at the linear order. From these expressions, we may find the 3-point functions treating the off-shell graviton $\Tilde{h}_{\mu\nu}$ as an independent field unconstrained by the gauge conditions. Using this prescription, one finds that each term of the Lagrangian (\ref{StelleSimplifiedAction}) gives rise to the following equations of motion at second order \cite{Dona:2015tra}:
\begin{equation*}
\left(\frac{\delta S_{\text{E}}}{\delta g_{\mu\nu}}\right)^{(2)} = \frac{1}{2\kappa^2}\left(\frac{1}{2}g^{\mu\nu}R-R^{\mu\nu}\right)^{(2)} = \frac{1}{2 \kappa^2}\left[ \frac{1}{2}\eta^{\mu\nu}R^{(2)} - (R^{\mu\nu})^{(2)} \right] .
\end{equation*}
\begin{equation*}
\begin{split}
\left(\frac{\delta S_0}{\delta g_{\mu\nu}}\right)^{(2)} &=  \frac{1}{2\kappa^2} \gamma_0\left[ 2 \sqrt{-g}\left( g^{\mu\nu}\nabla^2 - \nabla^{\mu}\nabla^{\nu} + \frac{1}{4}g^{\mu\nu}R - R^{\mu\nu}\right)R\right]^{(2)} \\ &= \frac{\gamma_0}{\kappa^2} (\eta^{\mu\nu}\Box - \partial^{\mu}\partial^{\nu})R^{(2)} \: .
\end{split}
\end{equation*}
\begin{equation*}
\begin{split}
\left(\frac{\delta S_2}{\delta g_{\mu\nu}}\right)^{(2)} & = \frac{\gamma_2}{2\kappa^2} \left[ \sqrt{-g}\left( \frac{1}{2}g^{\mu\nu}R^{\alpha\beta}R_{\alpha\beta} - 2R^{\mu\alpha}R^{\nu}_{\alpha}\right)\right]^{(2)} +\\ & + \frac{\gamma_2}{2\kappa^2}\left[ \sqrt{-g}(-g^{\mu\alpha}g^{\nu\beta}\nabla^2 - g^{\mu\nu}\nabla^{\alpha}\nabla^{\beta})R_{\alpha\beta}\right]^{(2)}+\\ & + \frac{\gamma_2}{2\kappa^2}\left[ \sqrt{-g}(g^{\mu\alpha}\nabla^{\beta}\nabla^{\nu} + g^{\nu\alpha}\nabla^{\beta}\nabla^{\mu})R_{\alpha\beta}\right]^{(2)} + \\ & = \frac{\gamma_2}{2 \kappa^2}(-\eta^{\mu\alpha}\eta^{\nu\beta}\Box - \eta^{\mu\nu}\partial^{\alpha}\partial^{\beta}+\eta^{\mu\alpha}\partial^{\beta}\partial^{\nu} + \eta^{\nu\beta}\partial^{\alpha}\partial^{\mu})R_{\alpha\beta}^{(2)} \: ,
\end{split}
\end{equation*}
where the Ricci tensor and the Ricci scalar at second order in the perturbation are given by
\begin{equation}
\begin{split}
     R^{(2)}_{\mu\nu} \:= \:& \frac{1}{2}\partial_{\mu}h^{\alpha\beta}\partial_{\nu}h_{\alpha\beta} + h^{\alpha\beta}(\partial_{\beta}\partial_{\alpha}h_{\mu\nu} + \partial_{\mu}\partial_{\nu}h_{\alpha\beta} - \partial_{\beta}\partial_{\alpha}h_{\nu\alpha} - \partial_{\beta}\partial_{\nu}h_{\mu\alpha})\\ & + \:
    \partial^{\beta}h_{\mu}^{\alpha}(\partial_{\beta}h_{\nu\alpha} - \partial_{\alpha}h_{\nu\beta}) \: ,  \\
    R^{(2)} \:= \:& - \partial_{\beta}h_{\alpha\sigma}\partial^{\sigma}h^{\alpha\beta} + \frac{3}{2}\partial_{\sigma}h_{\alpha\beta}\partial^{\sigma}h^{\alpha\beta} \: .
\end{split}
\end{equation}
Similarly, for the on-shell graviton interactions, we only need to reproduce the previous analysis but now expanding the equations of motion to fourth order \cite{Dona:2015tra}.

\subsubsection{4-graviton scattering amplitudes at tree-level}

\begin{figure}[H]
	\centering
	\begin{subfigure}[H]{0.49\textwidth}
		\centering
		 \includegraphics[width=0.67\textwidth]{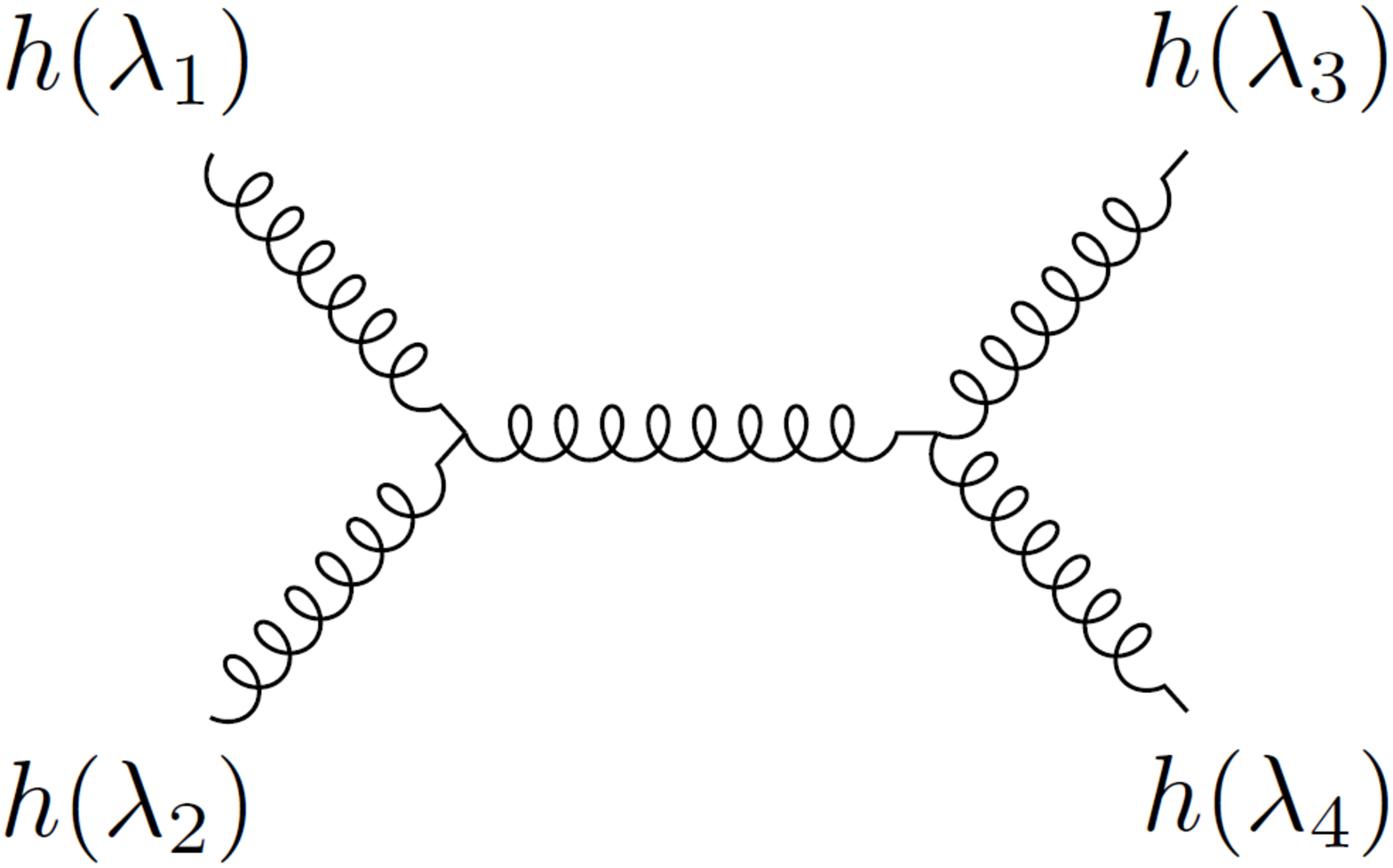}
		\caption{s channel}
	\end{subfigure}
	\begin{subfigure}[H]{0.49\textwidth}
		\centering
		\includegraphics[width=0.49\textwidth]{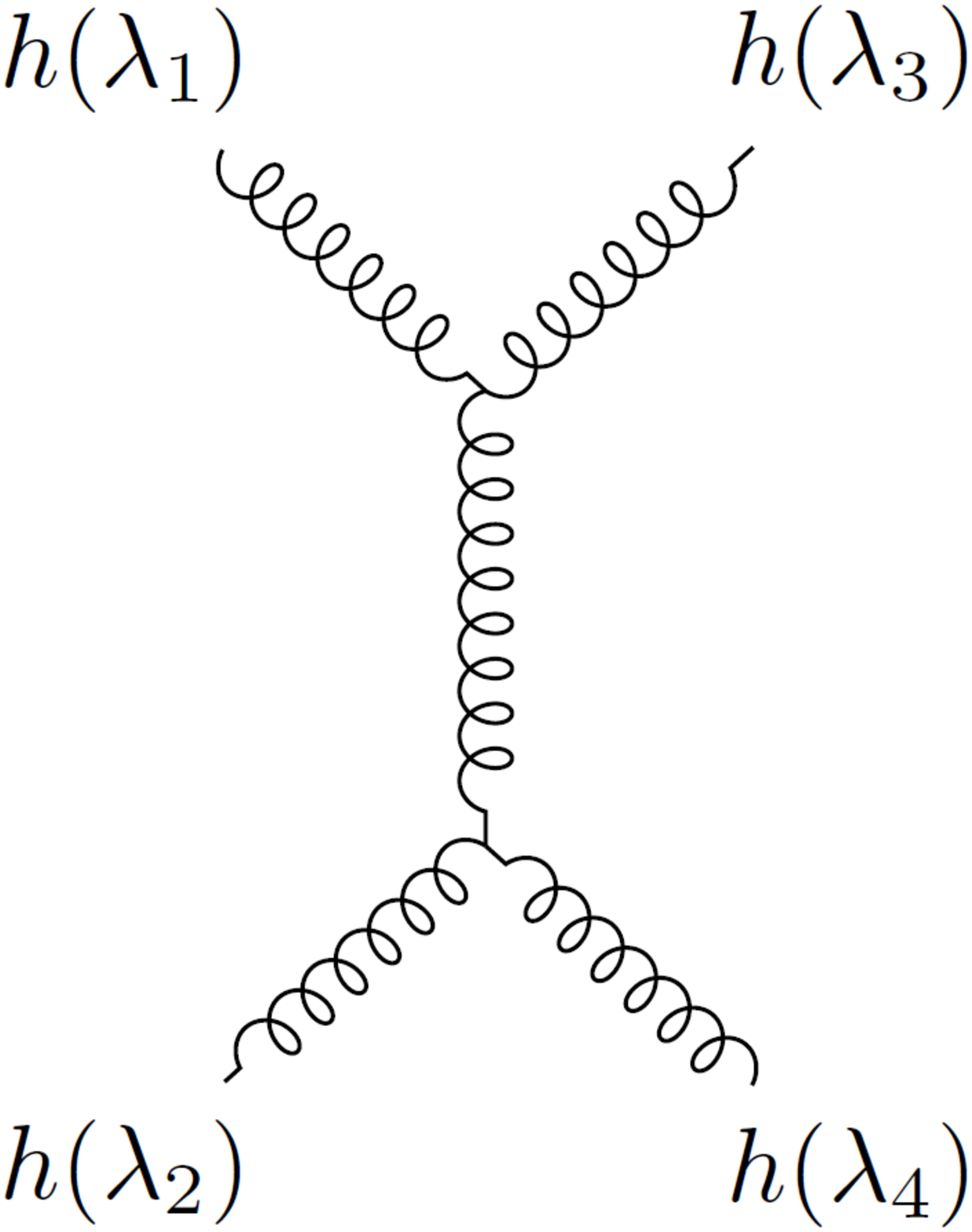}
		\caption{t channel}
	\end{subfigure}
	\begin{subfigure}[H]{0.49\textwidth}
		\centering
		\includegraphics[width=0.49\textwidth]{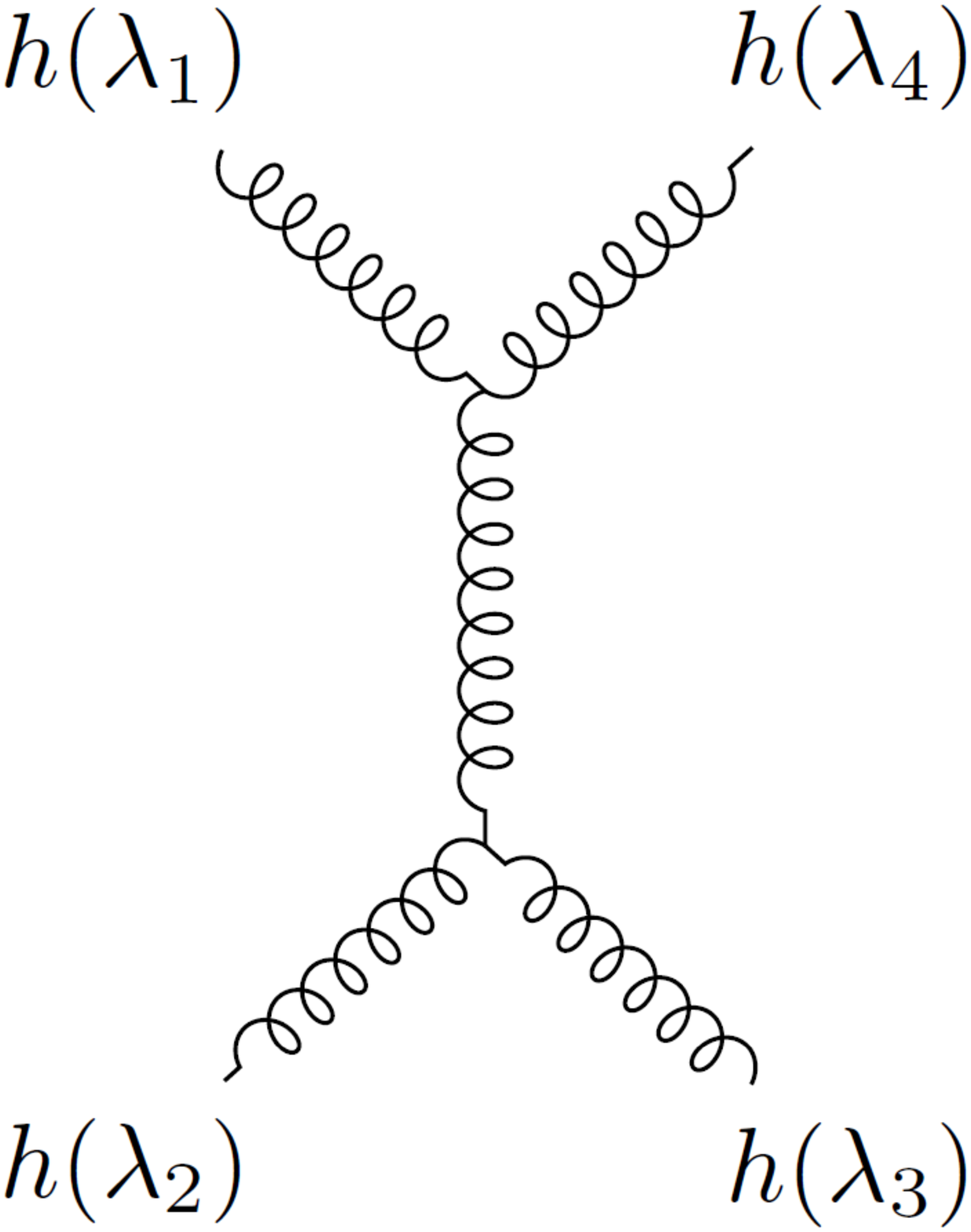}
		\caption{u channel}
	\end{subfigure}
	\begin{subfigure}[H]{0.49\textwidth}
		\centering
		\includegraphics[width=0.67\textwidth]{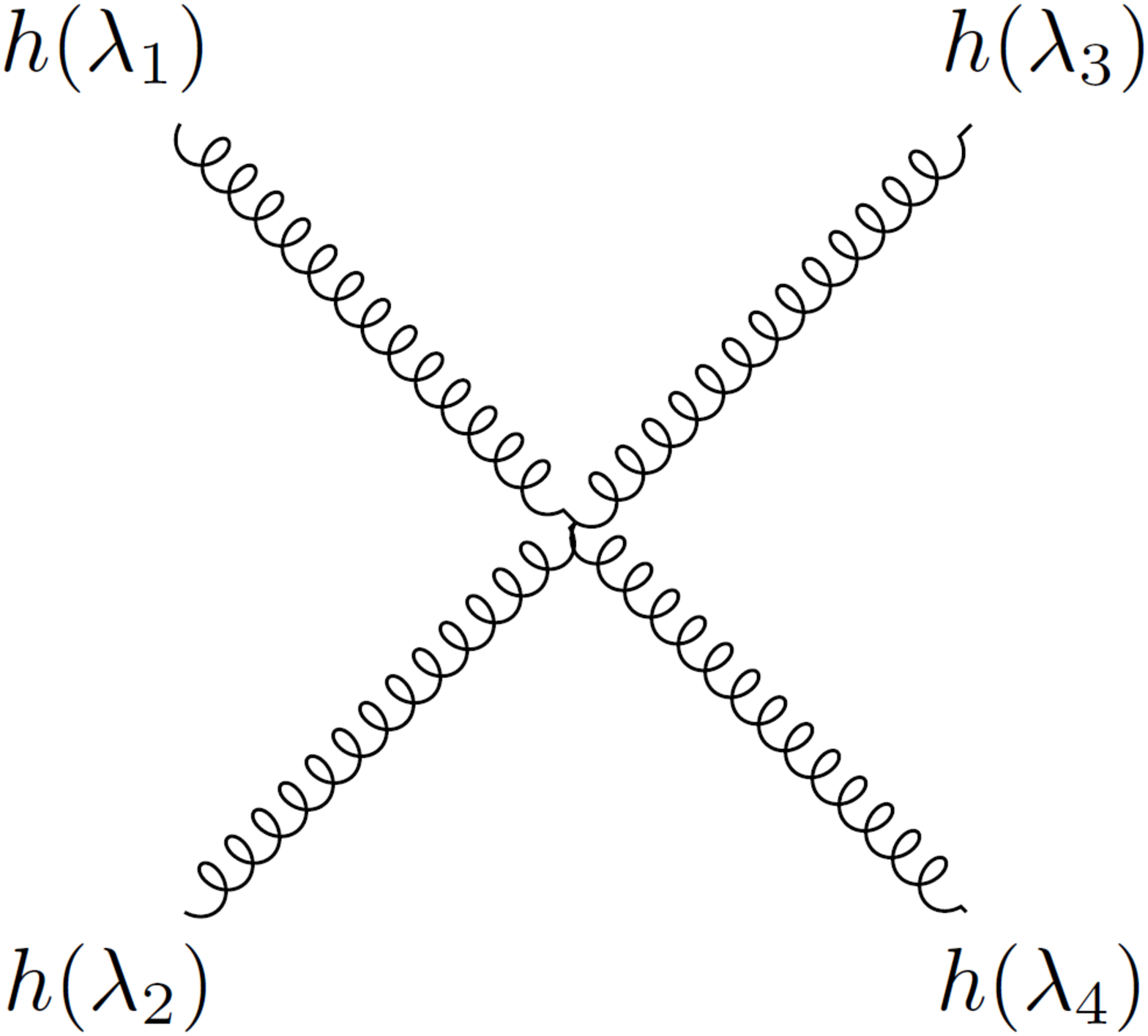}
		\caption{contact channel}
	\end{subfigure}
	\caption{The four contributions to the tree-level 2-2 graviton scattering.}
	\label{fig:4scattering}
\end{figure}

In this subsection, we recall the scattering amplitudes for the four available channels shown in Figure \ref{fig:4scattering} of the 2-2 graviton scattering, and to do so we make use of the Mandelstam variables for the process $p_1+p_2 \to p_3+p_4$, defined as
\begin{equation}
\begin{split}
     s&=-(p_1+p_2)^2 = 4E^2 \: ,\\
     t &= -(p_1-p_3)^2=-2E^2(1-\cos{\theta})\: ,\\
      u &= -(p_1-p_4)^2 = -2E^2(1+ \cos{\theta})\: ,
\end{split}
\end{equation}
where $E$ and $\theta$ are the energy and the scattering angle in the center-of-mass reference frame.
\noindent With all the necessary ingredients, one may proceed to calculate the four independent helicity amplitudes (\ref{4scatteringAmplitudes}). Adding the four contributions depicted in figure \ref{fig:4scattering}, one finds \cite{Dona:2015tra}
\begin{equation}
\begin{split}
    \mathcal{A}(++,++) \: = \: &  \frac{i}{\kappa^2}\frac{E^2}{\sin{\theta}^2} = \mathcal{A}_{\text{E}}(++,++) \: ,\\
    \mathcal{A}(++,+-)  \: = \: & 0 = \mathcal{A}_{\text{E}}(++,++)\: ,\\
     \mathcal{A}(+-,+-)  \: = \: &  \frac{1}{8} \frac{i}{\kappa^2}\frac{E^2(1+\cos{\theta})^4}{\sin{\theta}^2} = \mathcal{A}_{\text{E}}(+-,+-)\: ,\\
     \mathcal{A}(++,--) \: = \: & 0 = \mathcal{A}_{\text{E}}(++,++) \: . \label{eqStelleAmplitudes}
\end{split}
\end{equation}
The most impressive result is that, at the tree-level, the scattering amplitudes of Stelle's gravity coincide with the ones computed in Einstein's theory \cite{Berends:1974gk}.

\subsubsection{Helicity amplitudes in nonlocal gravity}

For the action (\ref{eqGenNLQG}), we may proceed analogously to what done in Stelle's gravity. In fact, since for the 4-graviton scattering amplitude we only need to compute the action at second and at fourth order in the perturbation $h_{\mu\nu}$ and since nonlocality modifies the terms proportional to $R_{\mu\nu}^2$ and $R^2$ and we have shown that these are at least quadratic in the perturbation, at the end of the day, the results obtained for the nonlocal case are the same than the higher-derivative theory. The nonlocal form factors will be mere spectators in the development because, at the order of the expansion we are interested in, they do not play any role so that the expressions of the scattering amplitudes remain intact and one recovers the values (\ref{eqStelleAmplitudes}) \cite{Dona:2015tra}. This result is a key to ensure causality.

\subsubsection{Causality}

Often, nonlocality has been associated with acausality, a property that rules out any theory by cause-effect violation; indeed, in the construction of a QFT via the second quantization, one requires causality in the sense that disconnected spacetime events cannot influence each other. However, this connection between acausality and nonlocality is not necessarily true and it can be shown that it does not hold in NLQG by reviewing the Shapiro time delay test \cite{Shapiro:1964uw}. 

Shapiro time delay is considered to be one of the four classical tests of GR, and it states that light propagating near a massive object will suffer a time delay with respect to the measured time while propagating in flat spacetime; this measurable delay is what is called Shapiro's time delay $\Delta t$ and it can be used to prove causality by showing that $\Delta t > 0$. A negative value of this measurable quantity would imply a violation of macro-causality.

To compute Shapiro's time delay one only needs to know the tree-level scattering amplitudes \cite{Kabat:1992tb,Ciafaloni:2014esa} and, as we have seen, for both Stelle's gravity and nonlocal quantum gravity one gets the same values of Einstein's theory. Since the causality of GR has been proven both theoretically and experimentally \cite{Shapiro:1971iv}, we conclude that NLQG does not induce macroscopic acausality. Microscopic but observationally harmless violations of causality are still possible \cite{Giaccari:2018nzr,Buoninfante:2018mre}.

\subsection{Renormalization} \label{SecRenormalization}

In this section, we study the renormalizability of NLQG by reviewing the superficial degree of divergence, as well as showing how multiplicative renormalization works in this theory. We also introduce the form of the so-called killer potential at least cubic in $\mathcal{R}$ that is required for the finiteness of the theory in $D=4$ dimensions. Let us start by commenting on the choice for the form factor, more precisely, $\text{H}(\Box)$. Although, the Wataghin and Krasnikov form factors are very insightful to show the behaviour of the form factors in the UV limit, in practice, one makes use of the asymptotically polynomial form factors, since their growth in the UV is well-defined by a polynomial $p(\Box)$ of degree $n_{\rm deg}$, so that the calculations become more straightforward. 

We focus on the minimal theory
\begin{equation}
    \mathcal{L} = \frac{1}{2\kappa^2}\left[ R + G_{\mu\nu} \frac{e^{\text{H}(\Box)}-1}{\Box} R^{\mu\nu} + \mathcal{V}(\mathcal{R})  \right]. \label{MinimalDLagrangian}
\end{equation}
It was shown \cite{Modesto:2014lga} that one may split (\ref{MinimalDLagrangian}) into its local and nonlocal part and that the nonlocal part of the Lagrangian does not contribute to the UV divergences of the theory but only to the finite part of its quantum effective action. Therefore, in order to analyze the infinities of the theory, one needs only to consider the local part, given by the UV limit of the previous Lagrangian, with $e^{\text{H}(\Box)} \sim p(\Box)$. Because of this argument, to study the power-counting renormalizability of the theory, one focuses on the local action
\begin{equation}
    S_{local} = \frac{1}{2\kappa^2}\int d^4x \sqrt{-g}\left[ R + \sum_{n=0}^{n_{\rm deg}-1} R a_n\Box^n R + \sum_{n=0}^{n_{\rm deg}-1} R_{\mu\nu} b_n\Box^n R^{\mu\nu}   \right] \label{SlocalPC} \: .
\end{equation}
where for the time being we have ignored $\mathcal{V}(\mathcal{R})$.

We can choose the polynomial of the action (\ref{SlocalPC}) to be a monomial of the highest order so that the UV behaviour is characterized by the dependence
\begin{equation*}
    \mathcal{R} \Box^{n_{\rm deg}-1}\mathcal{R} \: .
\end{equation*}
To perform an analysis about the superficial degree of divergence of the theory, we first need to account for the non-Abelian character of the gravitational theory. In this sense, in order to quantize via the path integral one needs to introduce three Faddeev--Popov ghost fields \cite{Buchbinder:1992rb}, which appear in the loops of the theory.

The superficial degree of divergence of the theory (\ref{SlocalPC}) is given by \cite{Modesto:2014lga}
\begin{equation}
    \omega(\mathcal{F})  = 4 - 2(n_{\rm deg}-1)(L-1) \: ,
\end{equation}
 and from this expression we notice that assuming 
\begin{equation}
    n_{\rm deg} > 3 \: ,
\end{equation}
only one-loop divergences survive, obtaining a super-renormalizable theory.

\subsubsection{Stelle's gravity}

The previous analysis can be generalized to other local theories such as Stelle's gravity in $D=4$ dimensions. For this theory, we have that $n_{\rm deg}=1$, so that at one loop
\begin{equation}
    \omega(\mathcal{F}) = 4 \: .
\end{equation}
However, all the apparent divergences of the theory can be reabsorbed by the quadratic operators appearing in the action, resulting in a successful renormalizable theory \cite{Stelle:1976gc}. This example shows how the power-counting argument is insufficient to fully determine the renormalization properties of a QFT.

\subsubsection{Finiteness}

One-loop divergences may be reabsorbed by the quadratic terms in the curvature in the Lagrangian, resulting in a renormalizable theory. Furthermore, it turns out that we can even define a finite quantum field theory by introducing a suitable potential, in which all the $\beta$-functions vanish. More precisely, one requires to introduce the \emph{killer operators} \cite{Modesto:2014lga}
\begin{equation}
    \mathcal{V}(\mathcal{R}) = s_R^{(1)}R_{\mu\nu}R^{\mu\nu}\Box^{n_{\rm deg}-3}R_{\rho\sigma}R^{\rho\sigma} + s_R^{(2)}R^2 \Box^{n_{\rm deg}-3} R^2 \: .
\end{equation}
The non-running coupling constants $ s_R^{(1)}$ and $ s_R^{(2)}$ only contribute linearly to the $\beta$-functions, so that one may always make a suitable choice to achieve finiteness, i.e., the absence of divergences at any loop order.

\subsubsection{Multiplicative renormalization}

Let us now sketch the multiplicative regularization using dimensional regularization prescription \cite{Peskin:1995ev}, in which one performs the renormalization procedure in a general dimension $d = 4 - \varepsilon$ and, subsequently, takes the limit $\varepsilon \to 0$ so that the divergent part of the theory is encoded in terms proportional to $1/\varepsilon$.
Afterwards, one reabsorbs these divergences with the counterterms induced by the multiplicative prescription, in which one redefines all the bare coupling constants with a perturbative expansion in terms of the renormalized coupling constants via
\begin{equation}
    \lambda_B = Z_{\lambda} \lambda_R \: .
\end{equation}
In this sense, after applying multiplicative renormalization one ends up with the renormalized Lagrangian
\begin{equation}
\begin{split}
    \mathcal{L}^R & \: = \: \mathcal{L} + \mathcal{L}_{\text{ct}}  \\
    & \: = \mathcal{L} + (Z_{\kappa^{-2}}-1)\frac{R}{2\kappa^2}  +  (Z_{a_0}-1)a_n R^2 + (Z_{b_0}-1)b_nR_{\mu\nu}R^{\mu\nu} \: , \label{L1}
\end{split}
\end{equation}
where in dimensional regularization one can write the Lagrangian of the 1-loop counterterms as a function of the $\beta$-functions of the coupling constants:
\begin{equation}
    \mathcal{L}_{\text{ct}} = \frac{1}{\varepsilon}\left[ \frac{1}{2}\beta_{\kappa^{-2}}R + \beta_{a_0}R^2 + \beta_{b_0}R_{\mu\nu}R^{\mu\nu} \right] \label{L2} \: .
\end{equation}
Since the local part of $\mathcal{L}$ contains the divergences of the theory in a set of terms we collectively denote as $\mathcal{L}_{\infty}$, they may be reabsorbed by setting
\begin{equation}
    \mathcal{L}_{\text{ct}} = - \mathcal{L}_{\infty} \: ,
\end{equation}
resulting in a renormalizable theory.

Comparing (\ref{L1}) and (\ref{L2}), one writes for the coupling constant $\alpha_i = \lbrace \kappa^{-2}, a_0, b_0\rbrace$
\begin{equation}
    (Z_{\alpha_i}-1)\alpha_i = \frac{1}{\varepsilon}\beta_{\alpha_i} \quad \implies \quad Z_{\alpha_i} = 1 + \frac{1}{\varepsilon}\beta_{\alpha_i}\frac{1}{\alpha_i} \: .
\end{equation}
However, since the vertices $\mathcal{R}^2$ do not give rise to divergences at one loop, one concludes that their $\beta$-functions are constant. In this way, we may solve analytically the renormalization group equations to find
\begin{equation}
    \beta_{\alpha_i} = \frac{1}{\mu}\frac{d \alpha_i}{d \mu} \quad \implies \quad \alpha_i(\mu) = \alpha_i(\mu_0) + \beta_{\alpha_i}\log{\frac{\mu}{\mu_0}}, \label{eqCouplings}
\end{equation}
where $\mu$ is the energy scale. From (\ref{eqCouplings}), one notices that all the three couplings exhibit the same running in energies.

\subsubsection{Asymptotic freedom} \label{secAF}

In (\ref{eqCouplings}), one notices the small growth of the coupling constants $\alpha_i$ with the energy scale $\mu$.  This hints at the property of asymptotic freedom: since the kinetic term of the theory grows higher in the UV limit than the couplings do, the interactions of NLQG become negligible in the UV. One can show mathematically such property for the minimal theory \cite{Asorey:1996hz}
\begin{equation}
\begin{split}
    S &\: = \:  \frac{1}{2 \kappa^2} \int d^4x \sqrt{-g} \left[ R + G_{\mu\nu}\frac{e^{\text{H}(\Box)}-1}{\Box}R^{\mu\nu} \right]\\
    & \: = \:\int d^4x \sqrt{-g}\left[ \omega R + \sum_{n=0}^{\infty}(a_nR\Box^n R + b_n R_{\mu\nu}\Box^nR^{\mu\nu}) \right] , \\
\end{split}
\end{equation}
where only $\omega (\mu)$, $a_0 (\mu)$, and $b_0 (\mu)$ have an energy dependence (\ref{eqCouplings}), and where $e^{\text{H}(\Box)}$ is assumed to be asymptotically a monomial for simplicity. Expanding this action as a function of the dimensionless graviton (\ref{Dimensionlessh}), one obtains \cite{Fradkin:1981iu} 
\begin{equation}
\begin{split}
    S & \: = \: \int d^4x \lbrace  \omega [h\Box h+h^2 \Box h + \mathcal{O}(h^4)] \\
     & \:\quad + \: \sum_ {n=0}^{\infty}a_n[ h \Box^{n+2}h + h^2 \Box^{n+2}h + \mathcal{O}(h^4) ] \\
    & \: \quad+ \: \sum_ {n=0}^{\infty}b_n[ h \Box^{n+2}h + h^2 \Box^{n+2}h + \mathcal{O}(h^4) ] \rbrace \label{ActionAF} \: ,
\end{split}
\end{equation}
where we have omitted the tensorial indices. A suitable rescaling of the graviton field given by
\begin{equation}
    h_{\mu\nu} \to \frac{1}{\sqrt{b_0(\mu)}}h_{\mu\nu} \equiv f(t) h_{\mu\nu}, \quad f^2 = \frac{f_0^2}{1 + f_0^2 \beta_b \log \frac{\mu}{\mu_0}} \label{rescalingG}\: ,
\end{equation}
allows us to infer the properties of the theory in the UV. Under this rescaling, $[h] = 1$ and the action (\ref{ActionAF}) becomes
\begin{equation}
\begin{split}
    S & \: = \: \int d^4x \lbrace \omega[f^2 h\Box h + f^3 h^2 \Box h + \mathcal{O}(f^4h^4)] \\
    & \:\quad + \: \sum_ {n=0}^{\infty}a_n[ f^2 h \Box^{n+2}h + f^3 h^2 \Box^{n+2}h + \mathcal{O}(f^4h^4) ] \\
    & \:\quad + \: \sum_ {n=0}^{\infty}b_n[ f^2 h \Box^{n+2}h + f^3 h^2 \Box^{n+2}h + \mathcal{O}(f^4h^4) ] \rbrace \label{ActionAF2} \: .
\end{split}
\end{equation}
From this expression, one notes that since $f \to 0$ when $\mu \to \infty$, the action at leading order involves only two gravitons, so that all interactions are suppressed in the UV limit. Furthermore, the rescaling (\ref{rescalingG}) ensures the validity of perturbation theory in the high-energy regime.

\section{Conclusions and prospects}

In this chapter, we have reviewed Stelle's gravity, a theory that successfully removes the quantum divergences of GR by the introduction of higher-derivative terms in the action. However, these terms lead to instabilities both at the classical and quantum level. Therefore, in order to solve this breaking of unitarity, we have motivated the introduction of nonlocal gravity. Preserving unitarity drives us to consider exponential and asymptotically polynomial nonlocal operators, whose construction is based on: (i) the recovery of GR at low energies, (ii) the desired behaviour of the propagator of the theory in the UV so that renormalizability and even finiteness of the theory are achieved, and finally (iii) the classical singularity problem of Einstein's theory may be overcome. Although the introduction of nonlocality in the theory may be deemed as problematic because it obscures the solution of the problem of initial conditions, the diffusion method provides a rigorous prescription to clarify these apparent problems.

We have explored the way one quantizes the nonlocal gravitational theory following standard techniques of perturbative field theory. The nonlocal components in the Lagrangian improve the convergence of the Feynman diagrams and avoid the introduction of new poles in the  graviton propagator, so that unitarity is preserved. Furthermore, we have shown that this theory exhibits asymptotic freedom, a property that allows one to neglect graviton interactions at extremely small distances. In particular, this property could lead to the avoidance of singularities in the classical regime.

In the near future, it would be useful to extend this analysis including matter content in the theory, since it is believed that all the astronomical black holes in the universe have been formed by a gravitational collapse. In this way, one would be able to characterize the matter content in the classical singularity and check if this singularity would be smoothened out by means of nonlocality. It should not be disregarded that, as explained in section \ref{secNLG}, the equations of motion derived from this theory are extremely complicated and, in general, exact solutions will be difficult to find unless we consider Ricci-flat spacetimes. A new formulation of NLQG based on a non-minimal coupling between gravity and matter fields overcomes these issues easily \cite{Modesto:2021ief,Modesto:2021okr,Modesto:2021soh}.

Whereas this nonlocal approach provides engaging answers at the theoretical level, quantum effects are expected to manifest at Planck energies, so that the experimental evidence of these theories will be difficult to measure. Nevertheless, the rich properties of the theory may be used to apply the nonlocal formalism to other branches of physics, such as condensed matter, particle physics and cosmology, where the energy scale of the processes are far below the Plank scale. Recent results in the new NLQG formulation with non-minimal matter coupling points towards observable consequences in the cosmology of gravitational waves \cite{Calcagni:2022tuz}.

In conclusion, we have reviewed the nonlocal formulation of quantum gravity with exponential and asymptotically polynomial operators, that succeeds in building a renormalizable and unitarity, or even completely UV-finite theory. Moreover, the use of nonlocal operators can provide a reasonable solution to the long-standing singularity problem. Although the empirical check of this class of theories may not be available yet, it could be imminent \cite{Calcagni:2022tuz} and their consistency and robustness signal a promising scope of applicability, possibly in other branches of physics other than quantum gravity at the Planck scale.


\section*{Acknowledgements}

G.C.\ is supported by grant PID2020-118159GB-C41 funded by MCIN/AEI/10.13039/501100011033.  A.B.\ is supported by contract CPI-22-291 funded by CIDEGENT/2020/020.

\bibliography{Bibliography}
\bibliographystyle{h-physrev}

\end{document}